\documentclass[]{article}
\usepackage[affil-it]{authblk}
\usepackage[a4paper, total={6.5in, 8in}]{geometry}
\usepackage{graphicx}
\usepackage{pgfplots}
\usepackage{booktabs} 
\usepackage{graphicx}
\usepackage{pgfplots}
\usepackage[all]{nowidow}
\usepackage[utf8]{inputenc}
\usepackage{tikz}
\usepackage{fancyhdr}
\usepackage{dblfloatfix}
\usepackage{float}
\usepackage{textgreek}
\usepackage{multicol}
\usepackage{algorithm}
\usepackage{relsize}
\usepackage{mathtools}

\usepackage{hyperref}
\usepackage{subfig}
\usepackage{verbatim}
\usepackage{algorithmic}
\usepackage[inline]{enumitem} 
\usepackage{setspace}

	\title{\textbf{Multinomial Cluster-Weighted Models for High-Dimensional Data}}

\author{\textbf{Kehinde Olobatuyi}}
\affil{Department of Statistics and Mathematical Finance,\\ University of Milano-Bicocca, Italy \\ k.olobatuyi@campus.unimib.it}

\author{\textbf{Oludare Ariyo}}
\affil{Department of Statistics, \\ Federal University of Agriculture, Abeokuta, Nigeria. \\ ariyoso@funaab.edu.ng}

\date{}

\definecolor{blue}{HTML}{1F77B4}
\definecolor{orange}{HTML}{FF7F0E}
\definecolor{green}{HTML}{2CA02C}

\pgfplotsset{compat=1.14}

\newcommand\MEETtitle[1]{\Large \bf \hskip2.25pc \parbox{.8\textwidth}{ \noindent%
		\large \bf \begin{center} #1 \end{center}\rm } \vskip.1in \rm\normalsize }

\newcommand\MEETauthor[1]{\hskip2.25pc \parbox{.8\textwidth}{ \noindent%
		\normalsize \bf \begin{center} #1 \end{center}\rm } \vskip-1pc }

\let\title\MEETtitle
\let\author\MEETauthor
\let\affil\MEETaddress

\begin{document}
	
	\maketitle
	
	
\begin{abstract}
Modeling of high-dimensional data is very important to categorize different classes. We develop a new mixture model called Multinomial cluster-weighted model (MCWM). We derive the identifiability of a general class of MCWM. We estimate the proposed model through Expectation-Maximization (EM) algorithm via an iteratively reweighted least squares (EM-IRLS) and Stochastic Gradient Descent (EM-SGD). Model selection is carried out using different information criteria. Various Adjusted Rand Indices are considered as a different measure of accuracy. The clustering performance of the proposed model is investigated using simulated and real datasets. MCWM shows excellent clustering results via performance measures such as Accuracy and Area under the ROC curve.		
		\medskip
		
		
	\end{abstract}	
	

\section{Background History}
	In model-based clustering and classification, cluster-weighted model (CWM) is a convenient approach when the random vector of interest consists of a response variable \textit{Y} and a set \textit{p} of explanatory variables \textit{X}. CWM is an input-output inference framework based on the probability density estimation of a joint set of input feature and output target data. A broad family of cluster weighted models assumes that the components of the conditional distributions belong to the exponential family and the covariates are either continuous or finite discrete or both (\cite{Ingrassiaetal2015}). Different types of exponential families such as Gaussian, Binomial, Poisson, and \textit{t} distributions have been considered for conditional distributions. \textit{t}-CWM is introduced by \cite{Ingrassiaetal2012}, and subsequently the twelve mixture models with random covariates are introduced in \cite{Ingrassiaetal2014}. A polynomial Gaussian CWMs to capture the non-elliptical data is introduced in \cite{Punzo2014}. The motivation of developing a family of CWMs is to solve the problem of assignment independence found in the finite mixture of linear regressions (\cite{Hennig2000}). However, the applicability of CWMs for multi-class in high-dimensional space still remains an open research problem.
	In this paper, we develop multinomial CWM to generally classify multi-class response data into different categories. We give an extensive derivation of the identifiability condition of the proposed model. To perform a maximum likelihood estimation of the proposed model, we derive the Expectation-Maximization algorithm from analytical and stochastic view such as iteratively reweighted least squares (IRLS) and Stochastic Gradient Descent (SGD) to make EM scalable for large dataset. For the first time we apply the variant of CWM to image classification problem, taking CWM from the perspective of regression to a classification.
	\par The paper is organized as follows. In Sect. \ref{sec:2}, we give a detailed explanation of the proposed model. The identifiability of the proposed model is derived in Sect. \ref{sec:3}. In Sect. \ref{sec:4}, we derive the estimation of the model parameters via EM algorithm. Furthermore, in Sect. \ref{sec:6} we present an experiment on simulated data and discuss convergence criterion. Sect. \ref{sec:7} shows the application of MCWMC on real datasets. Finally, in Sect. \ref{sec:8}, we draw some limitations and the main conclusions of this work.
\section{Model specification}\label{sec:2}
Let $(\mathbf{X}, \mathbf{Y})$ be a pair of random vector $\mathbf{X}$ of dimension $d$ and the multiclass response variable $\mathbf{Y}$ of dimension $J$ defined on $\mathcal{D}$ with a joint probability $p(\mathbf{x}, \mathbf{y})$. $\mathbf{X}$ takes values in some space $\mathcal{X} \subseteq \mathrm{R}^d$ and $\mathbf{Y}$ takes values in $\mathcal{Y} \subseteq \mathrm{R}^J$. Thus, $\b{x}, \b{y} \in \mathcal{X} \times \mathcal{Y} \subseteq \mathcal{R}^{d+J}$. Let $\mathcal{D}$ be partitioned into $G$ disjoint groups, say $\mathcal{D}_1,\cdot\cdot\cdot,\mathcal{D}_G$, i.e. $\mathcal{D} = \mathcal{D}_1 \cup \cdot\cdot\cdot \cup\mathcal{D}_G$. MCWM decomposes the joint probability $p(\mathbf{X},\mathbf{Y})$ as follows:
\begin{equation}
	p(\mathbf{X}, \mathbf{Y}) =  \sum_{g=1}^{G} p(\mathbf{Y}|\mathbf{X},\mathcal{D}_g) p(\mathbf{X}|\mathcal{D}_g) \pi_g,
	\label{equ:1}
\end{equation}
where $p(\mathbf{Y}|\mathbf{X},\mathcal{D}_g)$ is the conditional density of the multi-class response variable $\mathbf{Y}$ given the predictor vector $\mathbf{X}$ and $\mathcal{D}_g$, $p(\mathbf{X}|\mathcal{D}_g)$ is the probability density of $\mathbf{X}$ given $\mathcal{D}_g$, $\pi_g = p(\mathcal{D}_g)$ is the mixing weight of $\mathcal{D}_g$, with constraints $\pi_g>0$ and $\sum_{g=1}^{G}\pi_g = 1$, $g=1,...,G$.
\par We take the conditional density of multiclass response variable to be a Multinomial distribution $\mathbf{Y}|\mathbf{X},\mathcal{D}_g \sim Multi(\mathbf{\phi}_{jg}, ..., \mathbf{\phi}_{Jg})$, and the marginal density to be a Gaussian distribution $\mathbf{X}|\mathcal{D}_g \sim \mathcal{N}_d(\mathbf{\mu}_g,\mathbf{\Sigma}_g)$. Thereafter, we shall write $p(\mathbf{X}|\mathcal{D}_g) = \psi_d(\mathbf{X};\mathbf{\mu}_g, \mathbf{\Sigma}_g)$ and $p(\mathbf{Y}|\mathbf{X},\mathcal{D}_g) = p(\mathbf{Y};\mathbf{\Phi}_g)$, $g = 1,.., G$, where the conditional densities are based on the nonlinear mappings which we will define later. Thus, Equation \ref{equ:1} becomes:
\begin{equation}
	p(\mathbf{X},\mathbf{Y};\mathbf{\Theta}) = \mathlarger \sum_{g=1}^{G} \pi_g p(Y_1 = y_1,\cdot \cdot \cdot,Y_J = y_J|\mathbf{\phi}_{jg}, \cdot \cdot \cdot, \mathbf{\phi}_{Jg})\psi_d(\mathbf{X};\mathbf{\mu}_g,\mathbf{\Sigma}_g),
	\label{equ:3}
\end{equation}
where $\mathbf{\Theta} = (\mathbf{\Omega}, \vec{\mu}, \mathbf{\Sigma}, \vec{\pi})$ denotes the set of all model parameters, $\mathbf{\Phi} =(\mathbf{\phi}_{jg}, \cdot \cdot \cdot, \mathbf{\phi}_{Jg}) \in \mathrm{R}^{d \times J \times G}$ denotes the coefficient of the local model, location parameter $\vec{\mu} \in \mathrm{R}^{d \times G}$, and $\mathbf{\Sigma}$ is the positive definite covariance matrix.
Equation ~\eqref{equ:3} is referred to as Multinomial CWM and the posterior is given by
\begin{equation}
	p(\mathcal{D}_g|\mathbf{x},\mathbf{y}) = \frac{\pi_g p(Y_1 = y_1,...,Y_J = y_J|\mathbf{\phi}_{jg}, ..., \mathbf{\phi}_{Jg})\psi_d(\mathbf{x};\mathbf{\mu}_g,\mathbf{\Sigma}_g)}{\mathlarger \sum_{k=1}^{G} \pi_k p(Y_1 = y_1,...,Y_J = y_J|\mathbf{\phi}_{jk}, ..., \mathbf{\phi}_{Jk})\psi_d(\mathbf{x};\mathbf{\mu}_k,\mathbf{\Sigma}_k)}
	\label{equ:4}
\end{equation}
In Equation ~\eqref{equ:4}, $\psi(.)$ denotes the Gaussian density, the number of free parameters is $df = (GJd) + G[d(d+1)/2] + Gd + (G-1)$, and $G$ is the number of groups.
\par We relate the expected value of $\mathbf{Y}|\mathbf{\phi}_j$ to the covariates $\mathbf{X}$ through the relation $f_{jg}(\mathbf{x};\mathbf{\beta}_{jg}) = \mathbf{x}'_i\mathbf{\beta}_{jg}$, where
\begin{equation}
	f_{jg}(\mathbf{X};\mathbf{\beta}_{jg}) = \log \frac{\phi_{ijg}}{\phi_{i1g}} = \mathbf{X}'_i\mathbf{\beta}_{jg}.
	\label{equ:5}
\end{equation}
In Equation \ref{equ:5} $\beta_{0jg}$ is an intercept and $\mathbf{\beta}_{1jg}$ is a vector of regression coefficients, for $j = 2,...,J$. We note that the intercept $\beta_{0jg}$ is implicit in Equation \eqref{equ:5} and the model matrix $\mathbf{X}$ includes a column of ones.
Therefore, if $J = 2$ the MCWM reduces to Binomial CWM as a special case. The model used to predict the probabilities of different outcomes is given:
\begin{equation}
	\mathbf{\phi}_{jg} = \frac{\exp\{f_{jg}(\mathbf{x};\mathbf{\beta}_{jg})\}}{1 + \mathlarger \sum_{j=2}^{J} \exp\{f_{jg}(\mathbf{x};\mathbf{\beta}_{jg})\}}
	\label{equ:6}
\end{equation}
Equation \ref{equ:6} can be written as the original probabilities $\phi_{jg}$ rather than the log-odds by setting $\phi_{1g} = 0$. 
\subsection{Identifiability} \label{sec:3}
Sufficient identifiability conditions imply that any mixture distribution function from the specified model class can be uniquely parameterized, i.e. the parameters can be uniquely determined given infinitely many observations. On the contrary, if a mixture distribution is unidentifiable the parameters cannot be uniquely determined. Identifiability problems of finite mixture model can be categorized into \textit{trivial} and \textit{generic} problems \cite{Fruhwirth-Schnatter2006}. Trivial identifiability is a result of empty components with the same parameters and can be avoided by restraining the feasible parameter space $\omega$ to $\tilde{\omega}$ $\subset$ $\omega$, $\forall \hspace*{0.05in}\Theta \in \tilde{\omega}$: $\pi_g > 0, \hspace*{.05in} g = 1,...,G$.
\par \cite{Teicher1963}, \cite{Blischke1964}, and \cite{Titteringtonetal1985} argued that mixtures of binomial distributions with generic problem are identifiable if the condition $M \geq 2G - 1$ is fulfilled, where $M$ denotes the number of repetitions for a given individual. The restriction is necessary and sufficient for the model class of all mixtures with a maximum of $G$ components. The result obtained by \cite{Lindsay1995} for general class of mixtures of discrete exponential densities with $M + 1$ point supports the same condition applied for mixtures of multinomial distributions (\cite{GrunBettina2002}; \cite{ElmoreandWang2003}). Moreover, \cite{Hennig2000} showed that full rank covariance matrix is insufficient for identifiability of mixture of Gaussian regression models.
\par We establish the identifiability conditions and theorem for MCWM in order to estimate the parameters of model in Equation \eqref{equ:3}. Consider a parametric class of density function 
\begin{equation}
	\mathcal{F} = \{f(\mathbf{x};\psi):\mathbf{x} \in \mathcal{X}, \psi \in \Psi\}
	\label{equ:12}
\end{equation}
and the class of finite mixture of functions in $\mathcal{F}$, 
\begin{equation}
	\mathcal{H} = \biggl\{h(\mathbf{x};\mathbf{\varphi}):h(\mathbf{x};\mathbf{\varphi}) = \mathlarger \sum_{g=1}^{G}f(\mathbf{x};\vec{\Psi}_g)\pi_g, \hspace*{0.1in}\text{with} \hspace*{0.1in}\pi_g > 0 \hspace*{0.1in}\text{and}\hspace*{0.1in}\mathlarger \sum_{g=1}^{G}\pi_g = 1, \notag
\end{equation}
\begin{equation}
	f(.;\psi_g)\in\mathcal{F}, g = 1,\cdot\cdot\cdot,G, \mathbf{\Psi}_g \neq \mathbf{\Psi}_s \hspace*{0.05in} \text{for} \hspace*{0.05in} g \neq s, G \in \mathcal{N}, \mathbf{x} \in \mathcal{X}, \mathbf{\varphi} \in \mathbf{\Theta}\biggr\}
	\label{equ:13}
\end{equation}
This class is identifiable if for any two members of $\mathcal{H}$ such as
\begin{equation}
	h(\mathbf{x};\varphi) = \mathlarger\sum_{g=1}^{G}f(\mathbf{x};\mathbf{\psi}_g)\pi_g, \hspace*{0.1in} \text{and} \hspace{0.1in} h(\mathbf{x};\tilde{\mathbf{\varphi}}) = \mathlarger\sum_{v=1}^{\tilde{G}}f(\mathbf{x};\tilde{\mathbf{\psi}}_s)\tilde{\pi}_s.
	\label{equ:14}
\end{equation}
The equality $h(\mathbf{x};\mathbf{\varphi}) = h(\mathbf{x};\tilde{\mathbf{\varphi}})$ implies that $G = \tilde{G}$ and there exists a one-to-one correspondence between the two sets $\{1,...,G\}$ and $\{1,...,\tilde{G}\}$, such that $\pi_g = \tilde{\pi}_s$ and $\psi_g = \tilde{\psi}_s$. The class of the MCWM is given by 
\begin{equation}
	\mathcal{P} = \biggl\{p(\mathbf{x},\mathbf{y};\mathbf{\varphi}):p(\mathbf{x},\mathbf{y};\mathbf{\varphi}) = \mathlarger\sum_{g=1}^{G}F(\mathbf{y}|\mathbf{x};\mathbf{\beta}^g_j)p(\mathbf{x};\mathbf{\mu}_g, \mathbf{\Sigma}_g)\pi_g,\hspace{.1in} \text{with}\hspace{0.1in}\pi_g>0, \mathlarger\sum_{g=1}^{G}\pi_g=1,\notag
\end{equation}
\begin{equation}
	\mathbf{\beta}_j^g \neq \mathbf{\beta}_j^s \hspace*{0.1in}\text{for}\hspace*{0.1in} g \neq s, (\mathbf{x}',\mathbf{y})'\in \mathcal{R}\times\mathcal{Y}, \mathbf{\varphi}=\{\mathbf{\beta}_j^g, \mathbf{\mu}_g, \mathbf{\Sigma}_g, \pi_g; g = 1,...,G\}\in\mathbf{\Theta}, G \in \mathcal{N}\biggr\}
	\label{equ:15}
\end{equation}
We provide the sufficient conditions for $\mathcal{P}$ to be identifiable in $\mathcal{X}\times \mathcal{Y}$, where $\mathcal{X} \subseteq \mathcal{R}^d$ is a set with probability one according to Gaussian density $p(.)$.
\subsection*{Theorem} \textit{Let $\mathcal{P}$ be the class defined in Equation \eqref{equ:3} and assume there exists a set $\mathcal{X}\subseteq \mathcal{R}^d$ with probability one such that the mixture of multinomial distributions
	\begin{equation}
		\mathlarger\sum_{g=1}^{G} F(\mathbf{y}|\mathbf{x};\mathbf{\theta}^g)\gamma_g(\mathbf{x})
		\label{equ:16}
	\end{equation}
where $\log\bigg[\frac{\mathbf{\theta}^g_j}{\mathbf{\theta}^g_J}\bigg]=\mathbf{x}'_i\mathbf{\beta}_j^g$, is identifiable for each fixed $\mathbf{x} \in \mathcal{X}$, where $\gamma_1(\mathbf{x}), ..., \gamma_G(\mathbf{x})$ are positive weights summing to one for each $\vec{x} \in \mathcal{X}$. The class $\mathcal{P}$ is identifiable in $\mathcal{X} \times \mathcal{Y}$, if the following conditions are fulfilled:\\
for all $j = 1,...,J-1$ there exists a non-empty $\tilde{\mathbf{I}}_g$ which is a subset of $\cup_i \mathbf{I}_i$ and for which
\begin{equation}
	\mathlarger\sum_{i}\mathlarger\sum_{j} M_{ij} \geq 2G - 1 \hspace*{0.1in}\forall \hspace*{0.1in}i \in \tilde{\mathbf{I}}_g.
	\label{equ:18}
\end{equation}
where $\tilde{\mathbf{I}}_g$ is defined as the index set of all observations of individual $i$ with covariate vector $\mathbf{x}_i$.}
\par The condition in Equation \ref{equ:18} guarantees no intra-component label switching is possible. As the component membership is fixed for each individual, there exists a hyperplane that partitions the covariate points of the same group on one side of the divide. The condition implies there exists $i \in N$ with at least $2G - 1$ observations.
\subsection{The EM Algorithms for Parameter Estimation}\label{sec:4}
Let $(\mathbf{x}_1,\mathbf{y}_1),...,(\mathbf{x}_N,\mathbf{y}_N)$ be a sample drawn from model
\begin{equation}
	p(\mathbf{X},\mathbf{Y};\mathbf{\Theta}) = \mathlarger \sum_{g=1}^{G} \pi_g p(Y_1 = y_1,\cdot \cdot \cdot,Y_J = y_J|\mathbf{\phi}_{jg}, \cdot \cdot \cdot, \mathbf{\phi}_{Jg})\psi_d(\mathbf{X};\mathbf{\mu}_g,\mathbf{\Sigma}_g),
	\notag
\end{equation}
The corresponding likelihood, for a fixed number of components $G$, is given by 
\begin{equation}
	L(\mathbf{\Theta}) = \mathlarger \prod_{i=1}^{N}p(\mathbf{x}_i,\mathbf{y}_i;\big|\mathbf{\Theta}) = \mathlarger\prod_{i=1}^{N}\mathlarger\sum_{g=1}^{G} \pi_g p(\mathbf{y}_i|\mathbf{\phi}_{ig})\mathcal{N}(\mathbf{x}_i|\mathbf{\mu}_g,\mathbf{\Sigma_g})
	\label{equ:4.17}
\end{equation}
Define $z_i = (z_{i1},...,z_{iG})'$, with $z_{ig} = 1$ if $(\mathbf{x}_i',\mathbf{y}_i')'$ comes from $\mathcal{D}_g$, and $z_{ig} = 0$ otherwise, and consider the complete data $\big\{\big(\mathbf{x}_i',\mathbf{y}_i',\mathbf{z}_i'\big)'; i = 1,...,N\big\}$. Then the complete-data likelihood is as follows;
\begin{equation}
	L_{\mathcal{C}}(\mathbf{\Theta}) = \mathlarger \prod_{i=1}^{N}\mathlarger \prod_{g=1}^{G} \pi_g^{z_{ig}}p(\mathbf{y}_i|\mathbf{\phi}_{ig})^{z_{ig}}\mathcal{N}(\mathbf{x}_i|\vec{\mu}_g,\mathbf{\Sigma_g})^{z_{ig}}
	\label{equ:4.18}
\end{equation}
where $z_{ig}$ denotes the $gth$ component of $\mathbf{z}_i$. Taking the logarithm of Equation (\ref{equ:4.18}), we obtain
\begin{equation}
	l_\mathcal{C}(\mathbf{\Theta}) = \mathlarger\sum_{i=1}^{N}\mathlarger\sum_{g=1}^{G}\bigg[z_{ig}\ln p(\mathbf{y}_i|\mathbf{\phi}_{ig}) + z_{ig}\big \lbrace\ln\pi_g + \ln \mathcal{N}(\mathbf{x}_i|\mathbf{\mu}_g, \mathbf{\Sigma_g})\big\rbrace\bigg]
	\label{equ:4.19}
\end{equation}
\begin{equation}
	l_{1c}(\mathbf{\Omega}) = \mathlarger\sum_{i=1}^{N}\mathlarger\sum_{g=1}^{G}z_{ig}\ln \bigg\lbrace\prod_{j=1}^{J}\phi_{ijg}^{y_{ij}}\bigg\rbrace
	\label{equ:4.24}
\end{equation}
\begin{equation}
	l_{2c}(\mathbf{k}) = \mathlarger\sum_{i=1}^{N}\mathlarger\sum_{g=1}^{G}z_{ig}\ln\mathcal{N}(\mathbf{x}_i;\vec{\mu}_g,\mathbf{\Sigma}_g)
	\label{equ:4.27}
\end{equation}
\begin{equation}
	l_{3c}(\vec{\pi}) = \mathlarger\sum_{i=1}^{N}\mathlarger\sum_{g=1}^{G}z_{ig}\ln \pi_g
	\label{equ:4.28}
\end{equation}
where $\mathbf{\Omega}' = (\mathbf{\beta}_{01g}',...,\mathbf{\beta}_{1Jg}')'$, $\mathbf{k}=(\mathbf{k}'_1,...,\mathbf{k}'_G)'$ with $\mathbf{k}_g=(\mathbf{\mu}_g, \mathbf{\Sigma}_g)'$.
$Z$ is a matrix whose rows are vectors of ones and zeroes, one at the position of the group and zero everywhere else.
\par We use EM algorithm \cite{dempster1977etal} to optimize $l_{\mathcal{C}}(\mathbf{\Omega})$, $l_{\mathcal{C}}(\mathbf{k})$ and $l_{\mathcal{C}}(\mathbf{\pi})$ to find the maximum likelihood (ML) estimates for the unknown parameters of MCWM. The E-step on the $q$th iteration simply requires the calculation of the current conditional expectation of $Z_{ig}$ given the observed sample due to the linearity of $l_c(\tau)$ in the unobserved data $z_{ig}$, where $Z_{ig}$ is the random variable of $z_{ig}$. The conditional expectation of $Z_{ig}$ is given by:
\begin{equation}
	E_{\tau^{(q)}}(Z_{ig}|\mathbf{\Theta}) = z_{ig}^{(q)} = \frac{\bigg(\binom{M_i}{y_{i1},...,y_{iJ}}(\phi^{(q)})_{ijg}^{y_{ij}}...(\phi^{(q)})_{iJg}^{y_{iJ}}\bigg)\mathcal{N}\bigg(\mathbf{x}_i|\mathbf{\mu}^{(q)}_g,\mathbf{\Sigma}^{(q)}_g\bigg)\pi^{(q)}_g}{\mathlarger\sum_{k=1}^{G}\bigg(\binom{M_i}{y_{i1},...,y_{iJ}}(\phi^{(q)})_{ijk}^{y_{ij}}...(\phi^{(q)})_{iJk}^{y_{iJ}}\bigg)\mathcal{N}\bigg(\mathbf{x}_i|\mathbf{\mu}^{(q)}_k,\mathbf{\Sigma}_k^{(q)}\bigg)\pi_g^{(q)}} 
	\label{equ:4.30}
\end{equation} 
which corresponds to the posterior probability that the unobserved data $(x_i,y_i)'$ belong to the $gth$ component of the mixture, using the current fit $\vec{\tau}^{(q)}$ for $\vec{\tau}$ and $z_{ig}^{(q)}$ for $z_{ig}$ in Equation \eqref{equ:4.30}, we have 
\begin{equation}
	Q(\mathbf{\tau};\mathbf{\tau}^{(q)}) = Q_1(\mathbf{\Omega};\mathbf{\tau}^{(q)}) + Q_2(\mathbf{k};\mathbf{\tau}^{(q)}) + Q_3(\mathbf{\pi};\mathbf{\tau}^{(q)})
	\label{equ:4.31}
\end{equation} 
\par On the M-step, at the $(q+1)th$ iteration, we compute $\mathbf{\Omega}^{(q)}, \mathbf{k}^{(q)}$ and $\mathbf{\pi}^{(q)}$ in Equation (\eqref{equ:4.31}) independently of each other by separate maximization. $\mathbf{k}^{(q)}$ and $\mathbf{\pi}^{(q)}$ can be computed in closed form but $\mathbf{\Omega}^{(q)}$ has no closed-form solution.
\subsection*{Maximization of $Q_1(\mathbf{\Omega};\mathbf{\psi}^{(q)})$}
The updated estimates $\tau^{(q+1)}$ are the solutions of the derivatives.
\begin{equation}
	\frac{\partial}{\partial \mathbf{\beta}}Q_1(\mathbf{\Omega};\mathbf{\tau}^{(q)}) = \frac{\partial}{\partial \mathbf{\beta}}\mathlarger\sum_{i=1}^{N}\mathlarger\sum_{g=1}^{G}z^{(q)}_{ig} \Bigg\lbrace y_{i1}\ln\phi_{i1g} + \mathlarger\sum_{j=2}^{J}y_{ij}\ln\phi_{ijg}\Bigg\rbrace
	\label{equ:4.35}
\end{equation}
Since the derivative in Equation \eqref{equ:4.35} does not have a closed form, we resolve to iterative optimization which will be derived further in the supplementary. The updated estimates are 
\begin{equation}
	\beta^{(q+1)}_{jg} = \beta^{(q)}_{jg} + \Bigg(\mathlarger\sum_{i=1}^{N} z^{(q)}_{ig}n_i\mathbf{x}'_i v_{ijg}\mathbf{x}_i\Bigg)^{-1}\Bigg(\mathlarger\sum_{i=1}^{N} z^{(q)}_{ig}\mathbf{x}_i v_{ijg}\zeta^*_{ij}\Bigg)
	\label{equ:4.36}
\end{equation}
\begin{equation}
	\beta^{(q+1)}_{jg} = \Bigg(\mathlarger\sum_{i=1}^{N} z^{(q)}_{ig}n_i\mathbf{x}'_i v_{ijg}\mathbf{x}_i\Bigg)^{-1}\Bigg(\mathlarger\sum_{i=1}^{N} z^{(q)}_{ig}\mathbf{x}_i v_{ijg}\zeta^{(q)}_{ij}\Bigg)
	\label{equ:4.37}
\end{equation}
where $v_{ijg} = \phi^{(q)}_{ijg}(1-\phi^{(q)}_{ijg})$, $\zeta^{(q)}_{ij} = n_i\mathbf{x}_i\beta^{(q)}_{jg} +\zeta^*_{ij}$ and $\zeta^*_{ij} = y_{ij}/\phi^{(q)}_{ijg} - y_{i1}/\phi^{(q)}_{i1g}$. The weight $v_{ijg}$ and the adjusted response $\zeta^{(q)}_{ij}$ are updated at each iteration based on the current estimates of the multinomial distribution probability $\phi_{ijg}$.
\par We can also optimize using stochastic optimization if the cost function cannot be solved analytically. In this situation, we seek to minimize the negative log-likelihood by using the Stochastic Gradient Descent (SGD).
The gradients of the intercept and the coefficients are given by
\begin{equation}
	\beta_{0jg}^{(q+1)} = \beta^{(q)}_{0jg} - \alpha \frac{\partial}{\partial\beta_{0jg}}Q_1(\mathbf{\beta};\mathbf{\psi}^{(q)})
	\label{equ:330}
\end{equation}
\begin{equation}
	\vec{\beta}_{1jg}^{(q+1)} = \mathbf{\beta}^{(q)}_{1jg} - \alpha \frac{\partial}{\partial\mathbf{\beta}_{1jg}}Q_1(\mathbf{\beta};\mathbf{\psi}^{(q)})
	\label{equ:340}
\end{equation}
Learning rate $\alpha$ plays a vital role to tune the model.
\subsubsection*{\textit{Maximizing} $Q_2(\mathbf{k};\mathbf{\psi}^{(q)})$ \textit{by ML}}
With reference to the updated estimates of $\mathbf{k}_g$,
\begin{equation}
	Q_2(\mathbf{k};\mathbf{\psi}^{(q)}) = \mathlarger\sum_{i=1}^{N}\mathlarger\sum_{g=1}^{G}z^{(q)}_{ijg}\ln\psi_d(\mathbf{x}_i;\mathbf{\mu}_g,\mathbf{\Sigma}_g)
	\label{equ:800}
\end{equation}
we maximize equation ~\ref{equ:800} with respect to $\mathbf{\mu}_g$ and $\mathbf{\Sigma}_g$, $g = 1,\cdot\cdot\cdot, G$. The $(q+1)th$ estimate of $\vec{\mu}_g$ and the covariance matrix $\mathbf{\Sigma}_g$ and 
\begin{equation}
	\mathbf{\mu}_g^{(q+1)} = \mathlarger \sum_{i=1}^{N} z_{ig}^{(q)}\mathbf{x}_i\Bigg /\mathlarger \sum_{i=1}^{N}z_{ig}^{(q)}
	\label{equ:631}
\end{equation} 
\begin{equation}
	\mathbf{\Sigma}_g^{(q+1)} = \mathlarger \sum_{i=1}^{N} z_{ig}^{(q)}\big(\mathbf{x}_i - \mathbf{\mu}_g^{(q+1)}\big)\big(\mathbf{x}_i - \mathbf{\mu}_g^{(q+1)}\big)^T\Bigg /\mathlarger \sum_{i=1}^{N}z_{ig}^{(q)},
	\label{equ:640}
\end{equation} 
respectively.
\clearpage
\begin{table}[H]
	\centering\caption{True values of $\mathbf{\mu}$, $\mathbf{\sigma}^2$, and $\mathbf{\pi}$ for $G=2$ (top) and $G=3$ (bottom),\\ $N = 500, 1000$}\vspace{0.4cm}
	\label{tab:2}
	\begin{tabular}{r@{\hspace{.2in}}r@{\hspace{.2in}}r@{\hspace{.2in}}r@{\hspace{.2in}}r@{\hspace{.2in}}r@{\hspace{.2in}}rr}
		\hline
		&$g$&$\mu_1$&$\mu_2$&$\sigma^2_{11}$&$\sigma_{22}^2$&$\mathbf{\pi}$&\\ 
		\hline
		&$1$&$0.10$&$2.00$&$1.00$&$1.00$&$0.50$&\\
		&$2$&$-2.00$&$0.00$&$1.00$&$1.00$&$0.50$&\\
		\hline
		&$1$&$0.10$&$0.00$&$1.00$&$1.00$&$1/3$&\\ 
		&$2$&$-2.00$&$1.00$&$1.00$&$1.00$&$1/3$&\\ 
		&$3$&$2.00$&$3.00$&$1.00$&$1.00$&$1/3$&\\		
		\hline
	\end{tabular}
\end{table}
\bigskip
\subsection*{Maximizing $Q_3(\mathbf{\pi};\mathbf{\tau}^{(q)})$  by ML}
We maximize $Q_3(\mathbf{\pi};\mathbf{\tau}^{(q)})$ with respect to $\mathbf{\pi}$ subject to the parameter constraints using Lagrangian function
\begin{equation}
	\mathlarger\sum_{i=1}^{N}\mathlarger\sum_{g=1}^{G}z^{(q)}_{ig}\ln \pi_g - \lambda \bigg(\mathlarger\sum_{g=1}^{G}\pi_g-1\bigg)
	\label{equ:4.49}
\end{equation}
where $\lambda$ is the Lagrangian multiplier. The estimate of $\pi_g$ is 
\begin{equation}
	\pi_g^{(q+1)} = \frac{1}{N}\mathlarger \sum_{i=1}^{N} z_{ig}^{(q)}
	\label{equ:4.50}
\end{equation}
\section{A Simulation Study for Multinomial CWM}\label{sec:6}
We carried out a simulation study to evaluate the performance of the MCWM using EM algorithm. The data were simulated from Multinomial CWM according to the equation below
\begin{equation}
	p(\mathcal{D}_g|\mathbf{x},\mathbf{y}) = \frac{\pi_g p(Y_1 = y_1,...,Y_J = y_J|\mathbf{\phi}_{jg}, ..., \mathbf{\phi}_{Jg})\psi_d(\mathbf{x};\mathbf{\mu}_g,\mathbf{\Sigma}_g)}{\mathlarger \sum_{k=1}^{G} \pi_k p(Y_1 = y_1,...,Y_J = y_J|\mathbf{\phi}_{jk}, ..., \mathbf{\phi}_{Jk})\psi_d(\mathbf{x};\mathbf{\mu}_k,\mathbf{\Sigma}_k)}
	\notag
\end{equation}
We considered two cases $G = 2$ and $G = 3$ with two sample sizes of $N = 500$ and $N = 1000$ in the simulation. First, we initialized the parameters and set the dimension of covariates to two i.e., $d = 2$. The response variable $Y$ has $3$ categories $a, b$, and $c$ where $c$ is the baseline for both $a$ and $b$. Second, we generated two- and three-dimensional samples from each multivariate Gaussian distribution. The covariance matrices are identical matrices and different means were chosen for both cases $G = 2$ and $G = 3$ respectively. Table \ref{tab:2} (top) and Table \ref{tab:2} (bottom) present vectors of means and mixing proportions for cases $G = 2$ and $G = 3$. The response variable $Y$ defined as a multinomial variable with $3$ levels $(a,b$, and $c)$ was obtained from the probabilities in equation below 
\begin{equation}
	\mathbf{\phi}_{jg} = \frac{\exp\{f_{jg}(\mathbf{x};\vec{\beta}_{jg})\}}{1 + \mathlarger \sum_{j=2}^{J} \exp\{f_{jg}(\mathbf{x};\vec{\beta}_{jg})\}}.
	\notag
\end{equation}
Fixing the intercept $\vec{\beta}_{0g}$ at zero, the slopes $\vec{\beta}$ are presented in Table \ref{tab:3} (top) for $G = 2$ and Table \ref{tab:3} (bottom) for $G = 3$. We generated each data point according to the following setup: we sampled random numbers
\begin{figure}[H]
	\centering
	\begin{minipage}[b]{0.45\textwidth}
		\includegraphics[width = 3in, height=2in]{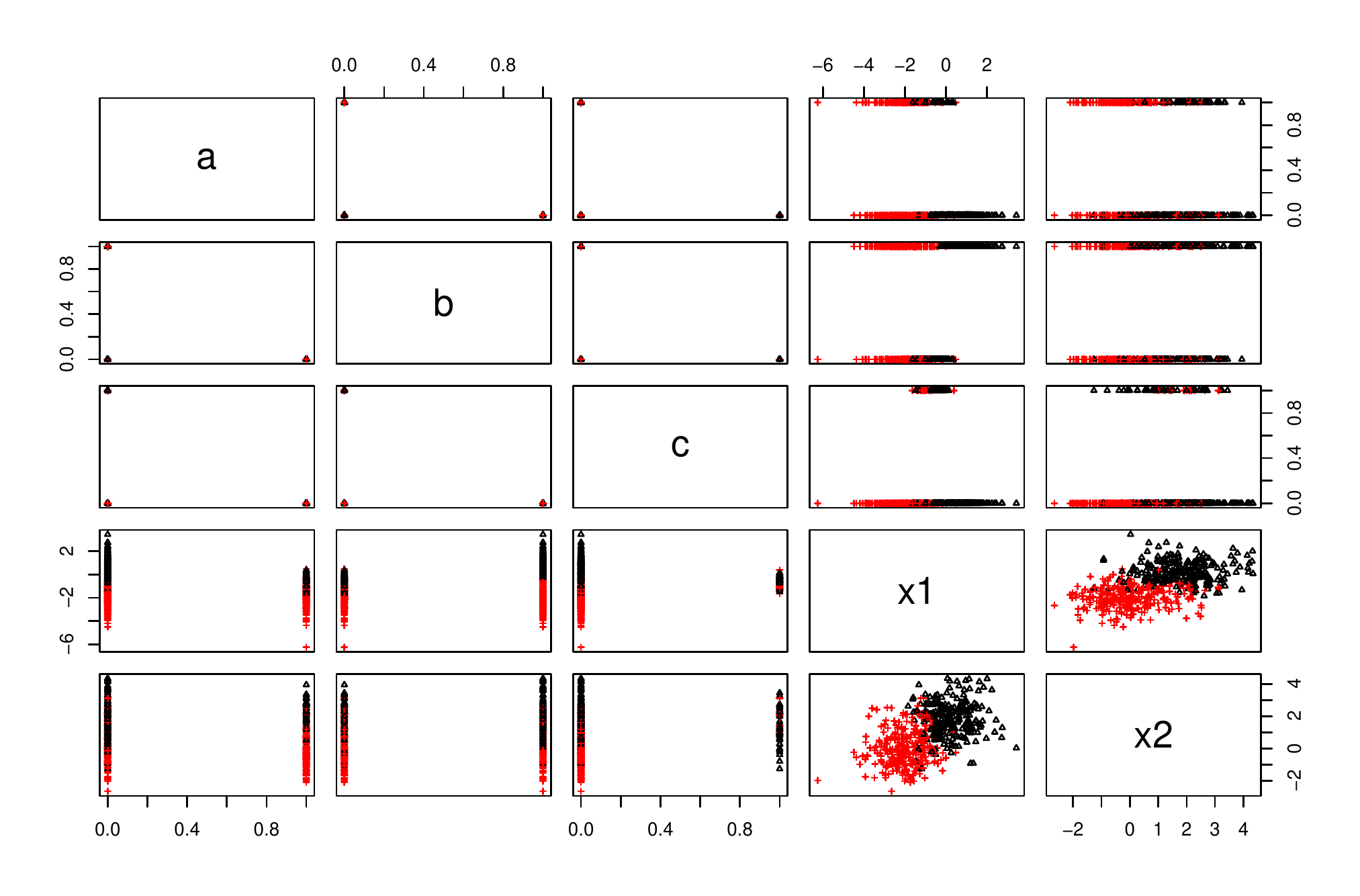}
		\centering\caption{\textit{The classification plot of multinomial CWM for ($N = 500$, $G = k = 2$) with two covariates.}} 
		\label{fig:1}
	\end{minipage}
	\hfil
	\begin{minipage}[b]{0.45\textwidth}
		\includegraphics[width = 3in, height=2in]{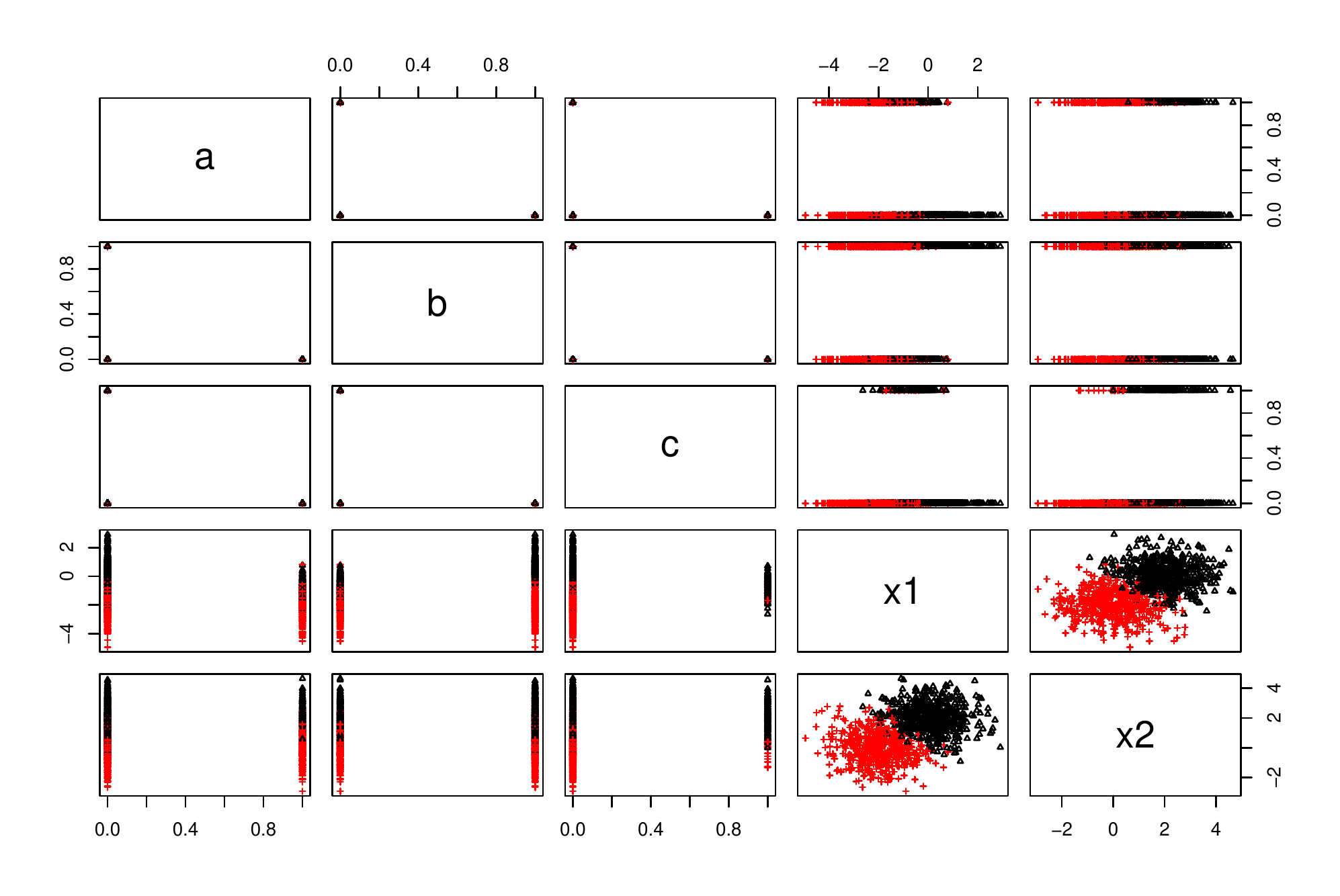}
		\centering\caption{\textit{The classification plot of multinomial CWM for ($N = 1000$, $G = k = 2$) with two covariates.}} 
		\label{fig:2}
	\end{minipage}
\end{figure}
\begin{table}[H]
	\centering\caption{True values of coefficients $\mathbf{\beta}$ for $N = 500, 1000$\\ and $G= 2, 3$}\vspace{0.4cm}
	\label{tab:3}
	\begin{tabular}{r@{\hspace{.3in}}r@{\hspace{.3in}}r@{\hspace{.3in}}r@{\hspace{.3in}}r@{\hspace{.3in}}r@{\hspace{.3in}}r}
		\hline
		&$g$ &$Y$&$\beta_0$&$\beta_1$& $\beta_2$&\\ 
		\hline
		&$1$&$a$&$0.000$&$5.000$&$0.400$&\\ 
		&&$b$&$0.000$&$0.300$&$0.040$&\\ 
		&$2$&$a$&$0.000$&$0.010$&$0.020$&\\
		&&$b$&$0.000$&$2.000$&$1.000$&\\
		\hline
		&$1$&$a$&$0.000$&$5.000$&$0.400$&\\ 
		&&$b$&$0.000$&$0.300$&$0.040$&\\ 
		&$2$&$a$&$0.000$&$0.010$&$0.020$&\\
		&&$b$&$0.000$&$2.000$&$1.000$&\\
		&$3$&$a$&$0.000$&$1.000$&$0.030$&\\
		&&$b$&$0.000$&$0.060$&$0.020$&\\		
		\hline
	\end{tabular}
\end{table}
\bigskip
\noindent of length $N$ from a Uniform distribution $U(0,1)$. We selected a particular component from MCWM using the generated values. We initialized component means $\mu_g$, mixing parameters $\pi_g$, and covariance matrices $\mathbf{\Sigma}_g$ according to Table \eqref{tab:2} and the coefficients according to Table \eqref{tab:3}. We generated $\mathbf{x}_i$ from Gaussian distribution with their respective group parameters $\vec{\mu}_g$ and $\mathbf{\Sigma}_g$. The probability $\phi_{i1}$ was computed using equation \eqref{equ:6}. We used eight information criteria for model selection of the true mixture components $G$. 
\par We used $\textsf{R}$ computing environment \cite{RDev2019} for the simulation. To provide a useful estimate of the limiting values, we adopted the Aitken's acceleration applicable to the sequence of the log-likelihood values ${l^{(p)}}$. \cite{Bohningetal1994} suggested EM algorithm can be stopped if $\big|l_\infty^{(p+1)} - l^{(p)}\big| < \epsilon$ for small $\epsilon$ and the sequence of the log-likelihood values is the main interest. However, we used the stopping rule $\epsilon = 0.05$ of \cite{McNicholas2010} and \cite{Punzo2014}.
\subsubsection{Discussion of two-component MCWM}
Figure \eqref{fig:1} and Figure \eqref{fig:2} show the plot partitioned by the classification result for $N = 500$ and $N = 1000$ with number of mixture components $G = 2$. 
Table \eqref{tab:4} shows the estimates of $\vec{\mu}$, $\vec{\pi}$ and the diagonal of $\mathbf{\Sigma}$. When $N = 500$, the estimates provided in group $2$ e.g, for $\hat{\mu}_1$ is $-1.944$ and for $N = 1000$, $\hat{\mu}_1$ is $-2.004$. The estimates for the coefficients are shown in Table \eqref{tab:5} with $c$ as the baseline. In Table \eqref{tab:6}, for $N = 500$, the 
\begin{figure}[H]
	\centering
	\begin{minipage}[b]{0.45\textwidth}
		\includegraphics[width = 3in, height=2in]{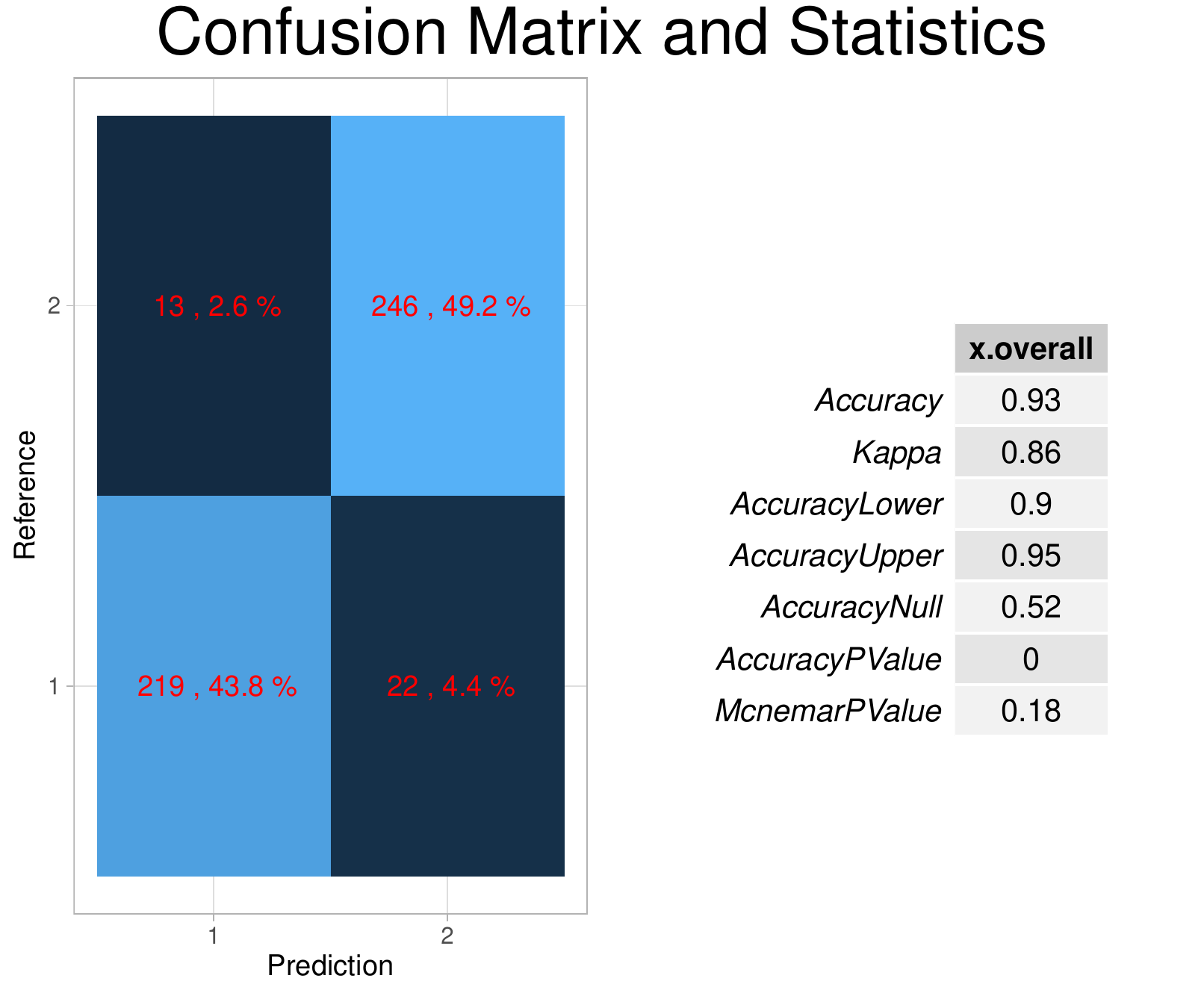}
		\centering\caption{\textit{The Visualization of Confusion Matrix and Statistics of the MCWM prediction with $N = 500$, $G = 2$.}} 
		\label{fig:3}
	\end{minipage}
	\hfil
	\begin{minipage}[b]{0.45\textwidth}
		\includegraphics[width = 3in, height=2in]{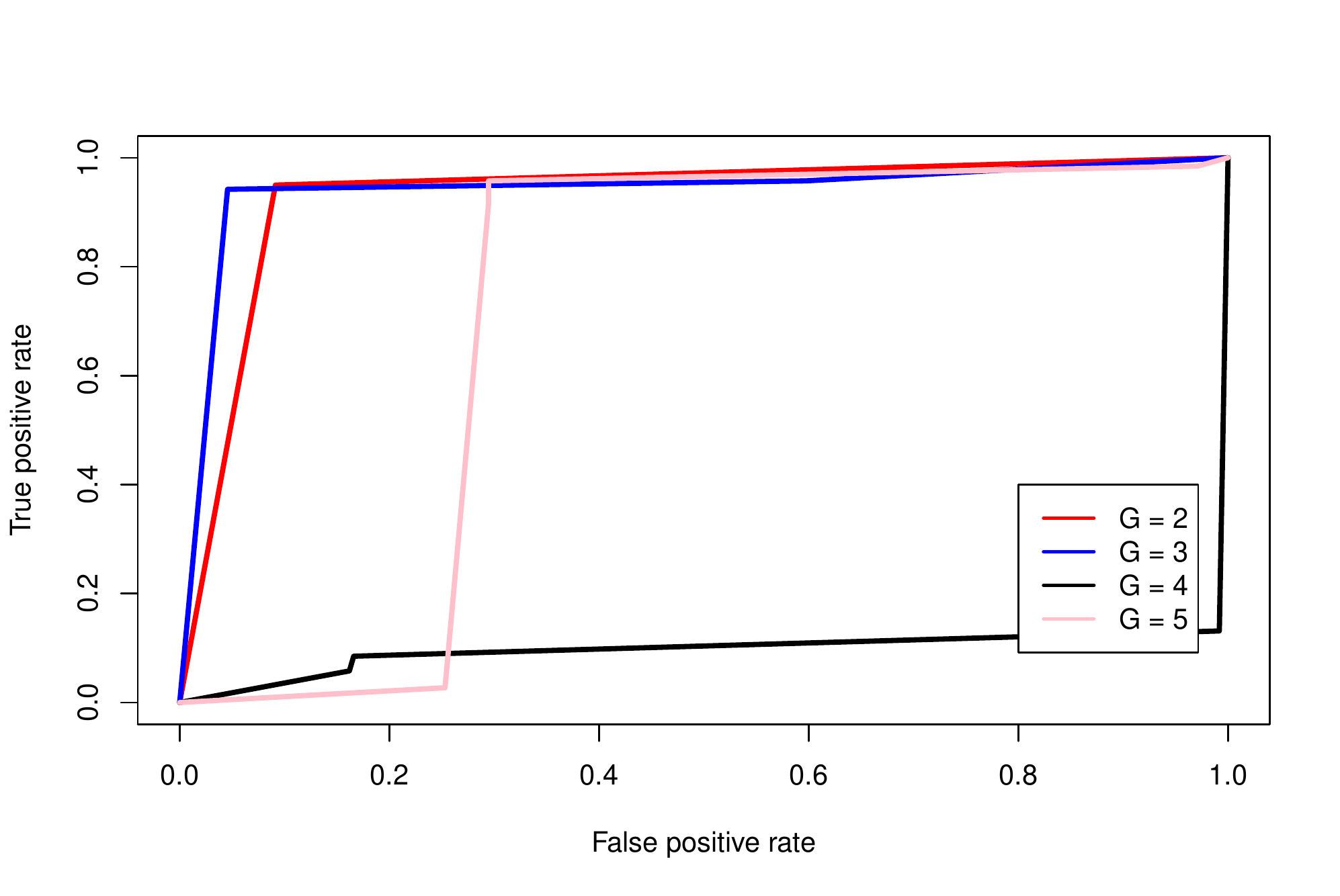}
		\caption{\textit{The Receiver's Operating Characteristic curve of the prediction by multinomialCWM ($N = 500$, $G = 2$).}} 
		\label{fig:4}
	\end{minipage}
\end{figure}
\bigskip
\begin{table}[H]
	\centering\caption{Estimated values of $\mathbf{\mu}$, $\mathbf{\sigma}^2$ and $\mathbf{\pi}$ for $N = 500, 1000$ and $G=2$}\vspace{0.4cm}
	\label{tab:4}
	\begin{tabular}{r@{\hspace{.2in}}r@{\hspace{.2in}}r@{\hspace{.2in}}r@{\hspace{.2in}}r@{\hspace{.2in}}r@{\hspace{.2in}}r@{\hspace{.2in}}r@{\hspace{.2in}}r}
		\hline
		&$N$&$g$&$\hat{\mu}_1$& $\hat{\mu}_2$&$\hat{\sigma}^2_{11}$&$\hat{\sigma}^2_{22}$&$\hat{\mathbf{\pi}}$&\\ 
		\hline
		True&&$1$&$0.100$&$2.000$&$1.000$&$1.000$&$0.500$&\\
		&&$2$&$-2.000$&$0.000$&$1.000$&$1.000$&$0.500$&\\
		\hline
		Recovered&$500$&$1$&$0.193$&$1.821$&$0.968$&$1.159$&$0.453$&\\
		&&$2$&$-1.944$&$0.046$&$0.946$&$1.245$&$0.547$&\\
		\hline
		Recovered&$1000$&$1$&$0.098$&$1.997$&$0.921$&$0.974$&$0.484$&\\
		&&$2$&$-2.004$&$-0.023$&$0.912$&$1.058$&$0.516$&\\
		\hline
	\end{tabular}
\end{table}
\begin{table}[H]
	\centering\caption{Estimated values of coefficients $\mathbf{\beta}$ for $N = 500, 1000$\\ and $G=2$ with $c$ as the baseline}\vspace{0.4cm}
	\label{tab:5}
	\begin{tabular}{r@{\hspace{.2in}}r@{\hspace{.2in}}r@{\hspace{.2in}}r@{\hspace{.2in}}r@{\hspace{.2in}}r@{\hspace{.2in}}rr}
		\hline
		&$N$&$g$&$Y$&$\hat{\beta}_0$&$\hat{\beta}_1$& $\hat{\beta}_2$&\\ 
		\hline
		True&&$1$&$a$&$0.000$&$5.000$&$0.400$&\\ 
		&&&$b$&$0.000$&$0.300$&$0.040$&\\ 
		&&$2$&$a$&$0.000$&$0.010$&$0.020$&\\
		&&&$b$&$0.000$&$2.000$&$1.000$&\\
		\hline
		Recovered&$500$&$1$&$a$&$0.298$&$6.889$&$0.159$&\\
		&&&$b$&$0.593$&$0.377$&$-0.489$&\\
		&&$2$&$a$&$0.348$&$0.143$&$0.156$&\\
		&&&$b$&$0.602$&$3.793$&$2.597$&\\
		\hline
		Recovered&$1000$&$1$&$a$&$0.378$&$4.887$&$0.183$&\\
		&&&$b$&$-0.118$&$0.160$&$-0.037$&\\
		&&$2$&$a$&$0.117$&$-0.004$&$-0.048$&\\
		&&&$b$&$-0.133$&$1.721$&$0.608$&\\
		\hline
	\end{tabular}
\end{table}
\begin{figure}[H]
	\centering
	\begin{minipage}[b]{0.45\textwidth}
		\includegraphics[width = 3in, height=2in]{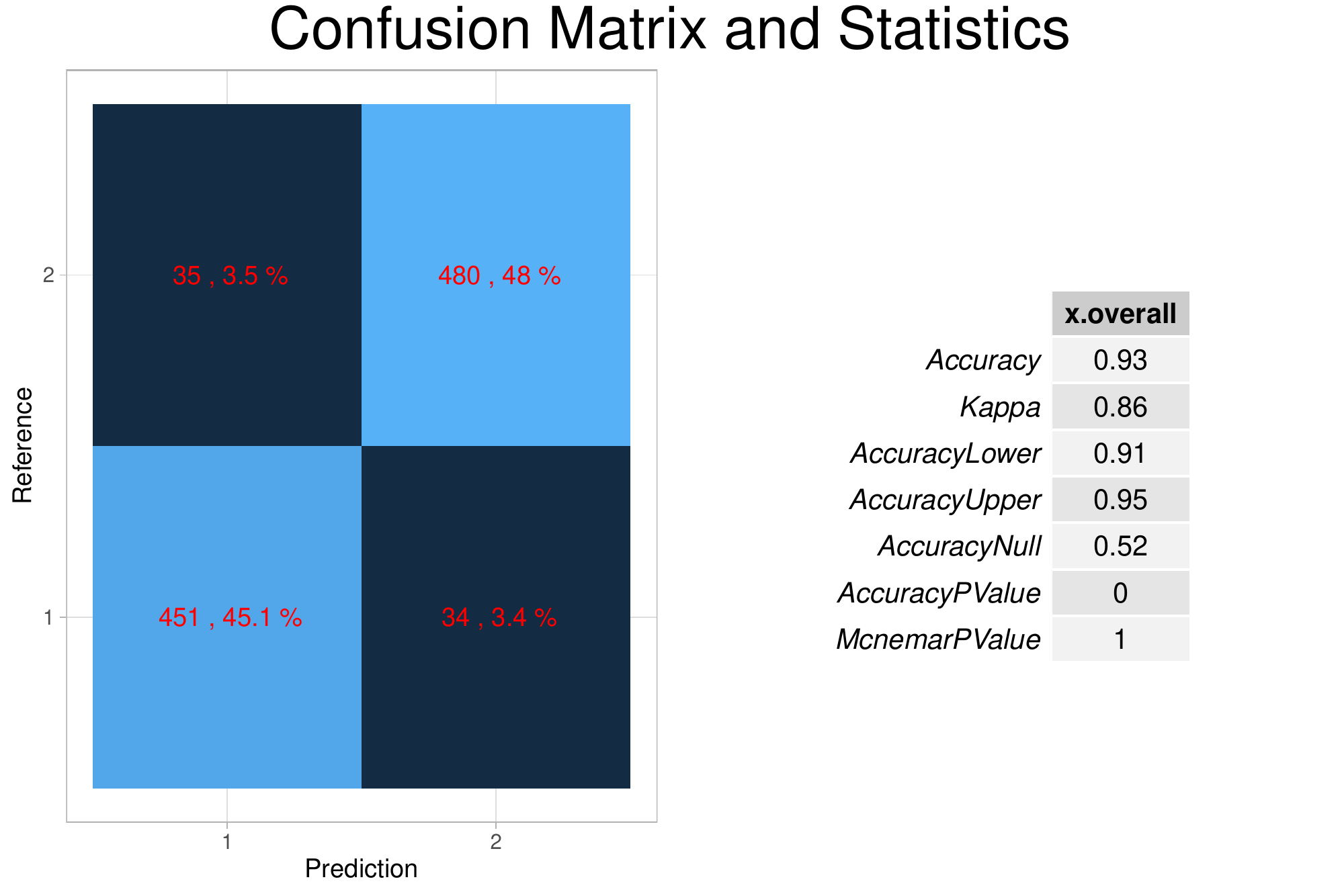}
		\centering\caption{\textit{The Visualization of Confusion Matrix and Statistics of the MCWM prediction with $N = 1000$, $G = 2$.}} 
		\label{fig:5}
	\end{minipage}
	\hfil
	\begin{minipage}[b]{0.45\textwidth}
		\includegraphics[width = 3in, height=2in]{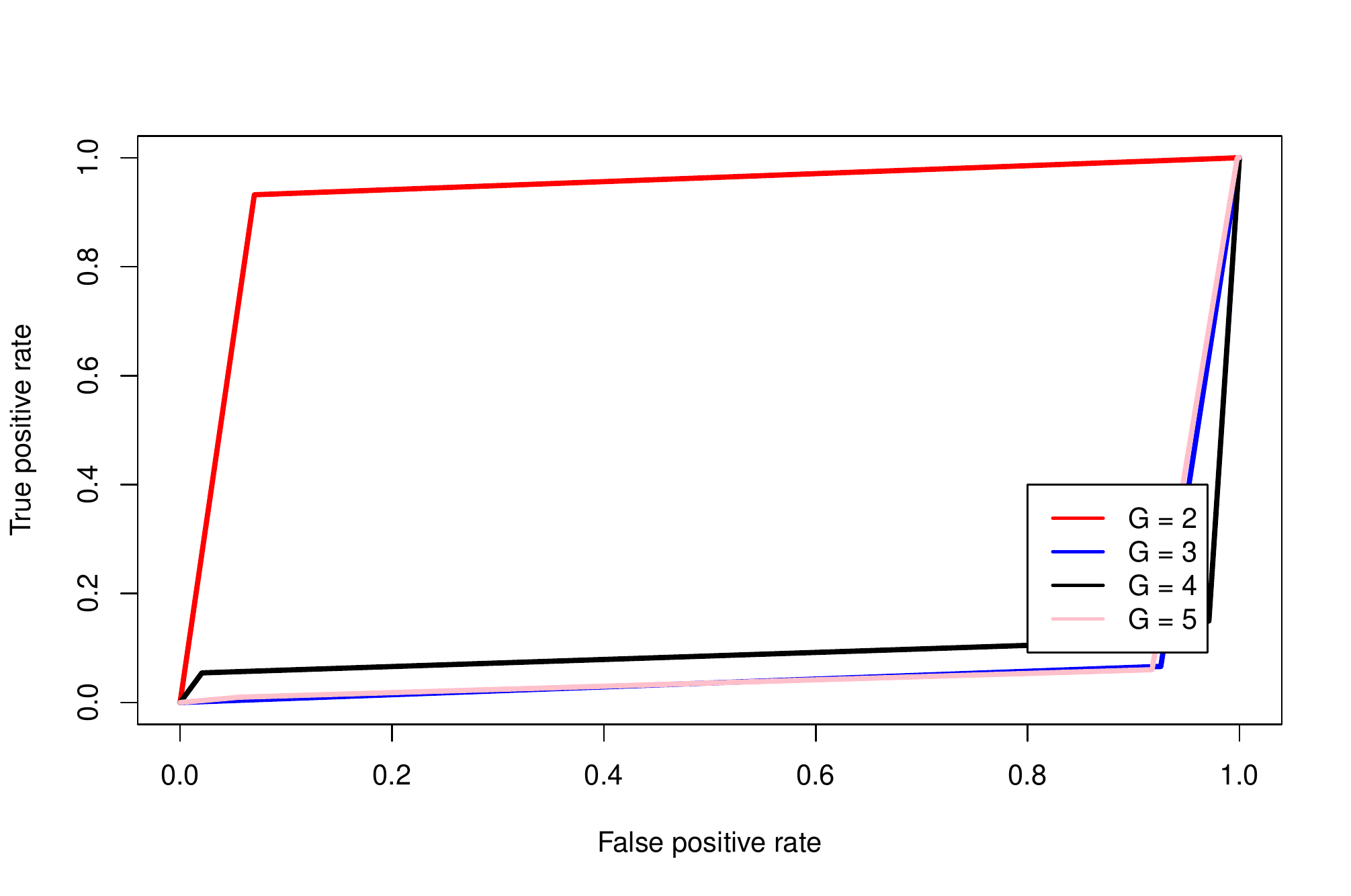}
		\caption{\textit{The Receiver's Operating Characteristic curve of the prediction by multinomialCWM ($N = 1000$, $G = 2$).}} 
		\label{fig:6}
	\end{minipage}
\end{figure}
\bigskip
\begin{table}[H]
	\centering\caption{Confusion Matrix in the three component Model for $N =500$ and $1000$}\vspace{0.4cm}
	\label{tab:6}
	\begin{tabular}{r@{\hspace*{.2in}}r@{\hspace*{.1in}}r@{\hspace*{.1in}}r@{\hspace*{.1in}}r@{\hspace*{.1in}}rrr@{\hspace*{.1in}}r@{\hspace*{.1in}}r@{\hspace*{.1in}}r@{\hspace*{.1in}}}
		\hline
		$G$&Real Component&&&$1$&&$2$&&&Misclassification rate (\%)&\\ 
		\hline
		$500$&$1$&&&$219$&&$13$&&&$5.60$&\\ 
		&$2$&&&$22$&&$246$&&&$8.21$&\\
		\hline
		rate&&&$7.00$&&&&&&&\\
		Accuracy&&&$93.00\%$&&&&&&&\\
		\hline
		$1000$&$1$&&&$451$&&$35$&&&$7.20$&\\
		&$2$&&&$34$&&$480$&&&$6.61$&\\
		\hline
		Misclassification&&&$6.90$&&&&&&&\\
		Accuracy&&&$93.10\%$&&&&&&&\\
		\hline
	\end{tabular}
\end{table}
\bigskip
\noindent model achieves an accuracy of $93\%$. In group one, the model has a misclassification rate of $5.60\%$, while in group two the model has a misclassification rate of $8.21\%$. The overall misclassification rate is $7.00\%$. The accuracy of the model when the sample size $N = 1000$ is $93.10\%$. However, the misclassification rate in group two is $7.20\%$, while the overall misclassification rate is $6.90\%$ (Table \ref{tab:6}). Figure \eqref{fig:3} and Figure \eqref{fig:4} show the visualization of the confusion matrix and Area under ROC curve for $N = 500$ respectively.
\par \noindent Similarly, Figure \eqref{fig:5} and Figure \eqref{fig:6} show the confusion matrix plot and Area under ROC curve of the prediction produced by multinomial CWM for $1000$. In Figure \ref{fig:4}, two-component MCMW $G = 2$ achieves higher accuracy than $G = 3,4,5$ considered. The area under ROC curve for $G = 2$ coincides with the area under the curve for the component $G = 3$, this might indicate that many of the classes are distributed across the two groups while only a few observations are clustered in the third group. Selection of three- or more component models may seem to be a consequence of spurious clusters. Spurious clusters are very common in the mixture modeling literature (\cite{Ingrassia2004}; \cite{IngrassiaandRocci2007}) and are commonly associated with low variance of a mixture component \cite{McandP2000}; \cite{IngrassiaandRocci2007}). However, such cases have been dealt with by placing bounds on the variance (\cite{Ingrassia2004}; Browne et al., 2013). This can be a future investigation of our work. Four-component MCWM $G = 4$ falls below $50\%$ accuracy which indicates an inappropriate model for the data, and the area under ROC curve of five-component $G = 5$ is lower than both $G = 2$ and $G = 3$. Similarly, in Figure \eqref{fig:6}, the components other than two components $G = 2$ perform randomly by guessing true positive and false positive less than $50\%$ of the time. The ROC 
\begin{table}[H]
	\centering\caption{The values of $8$ core selected information criteria of MCWM for different $N = 500,1000$ and $G=2$. The values in bold face are the smallest values selecting the true values of the artificial data.}\vspace{0.4cm}
	\label{tab:7}
	\begin{tabular}{r@{\hspace{.3in}}r@{\hspace{.1in}}r@{\hspace{.1in}}r@{\hspace{.1in}}r@{\hspace{.1in}}r@{\hspace{.1in}}r@{\hspace{.1in}}r@{\hspace{.1in}}r@{\hspace{.1in}}r@{\hspace{.1in}}r@{\hspace{.1in}}}
		\hline
		$N$&$G$&AIC&BIC&ICL&AWE&AIC3&AICc&AICu&Caic&\\ 
		\hline
		$500$&$\mathbf{2}$&$\mathbf{3641.63}$&$\mathbf{3742.78}$&$\mathbf{3739.40}$&$\mathbf{3223.33}$&$\mathbf{3521.63}$&$\mathbf{3639.11}$&$\mathbf{3613.46}$&$\mathbf{3420.48}$&\\
		&$3$&$3982.60$&$4134.32$&$4130.97$&$3355.15$&$3802.60$&$3976.84$&$3938.40$&$3650.87$&\\
		&$4$&$4633.17$&$4835.47$&$4831.39$&$3796.56$&$4393.17$&$4622.73$&$4571.16$&$4190.86$&\\
		&$5$&$4744.61$&$4997.48$&$4992.71$&$3698.85$&$4444.61$&$4727.93$&$4662.88$&$4191.73$&\\
		\hline
		$1000$&$\mathbf{2}$&$\mathbf{7031.83}$&$\mathbf{7149.62}$&$\mathbf{7140.55}$&$\mathbf{6580.26}$&$\mathbf{6911.83}$&$\mathbf{7030.60}$&$\mathbf{7005.28}$&$\mathbf{6794.05}$&\\
		&$3$&$9414.11$&$9590.79$&$9580.34$&$8736.75$&$9234.11$&$9411.34$&$9373.64$&$9057.43$&\\
		&$4$&$8983.30$&$9218.87$&$9209.89$&$8080.15$&$8743.30$&$8978.35$&$8928.11$&$8507.73$&\\
		&$5$&$11155.14$&$11449.60$&$11442.99$&$10026.21$&$10855.14$&$11147.34$&$11084.4$&$10560.67$&
		\\
		\hline
	\end{tabular}
\end{table}

\begin{table}[H]
	\centering\caption{Adjustment Rand Index and its variants of the three-component Model for $N=500$ and $1000$}\vspace{0.4cm}
	\label{tab:8}
	\begin{tabular}{r@{\hspace*{.3in}}r@{\hspace*{.3in}}r@{\hspace*{.3in}}r@{\hspace*{.3in}}r@{\hspace*{.3in}}r@{\hspace*{.3in}}r@{\hspace*{.3in}}rr}
		\hline
		$N$&$G$&Rand&HA&MA&FM&Jaccard&AUC&\\ 
		\hline
		$500$&$2$&$\mathbf{0.870}$&$\mathbf{0.739}$&$\mathbf{0.739}$&$\mathbf{0.870}$&$\mathbf{0.769}$&$\mathbf{0.930}$&\\
		&$3$&$0.812$&$0.624$&$0.625$&$0.794$&$0.648$&$0.939$&\\ 
		&$4$&$0.840$&$0.680$&$0.681$&$0.827$&$0.697$&$0.901$&\\
		&$5$&$0.802$&$0.604$&$0.605$&$0.781$&$0.627$&$0.709$&\\
		\hline
		$1000$&$2$&$\mathbf{0.871}$&$\mathbf{0.743}$&$\mathbf{0.743}$&$\mathbf{0.871}$&$\mathbf{0.772}$&$\mathbf{0.931}$&\\
		&$3$&$0.868$&$0.736$&$0.736$&$0.868$&$0.766$&$0.929$&\\
		&$4$&$0.824$&$0.648$&$0.649$&$0.809$&$0.673$&$0.900$&\\
		&$5$&$0.851$&$0.703$&$0.703$&$0.847$&$0.735$&$0.925$&\\
		\hline
	\end{tabular}
\end{table}
\bigskip
\noindent plot in Figure \eqref{fig:6} gets improved with large sample size $N = 1000$. There is a clear-cut distinction between area under the ROC curve of two-component and three-component MCWM with large datasets.  
\par \noindent Table \eqref{tab:7} shows the values of information criteria. We compared different groups to investigate the identifiability power of the model. We observed that all the eight criteria agree with the two-component model which is the true component of the model. The selection criteria of $G = 2$ for both sample sizes have the smallest values that shows the model is identifiable for simulated data. Table \eqref{tab:8} shows the values for the ARI and its variants to further establish the selection of the true component and performance evaluation of the model. The higher the values, the stronger the agreement between the actual classes and the predicted classes. The two-component MCWM has the highest values among other number of components in both sample sizes. This simply means that the model with the components other than the true component performed poorly in this simulation study.
\subsubsection{Discussion of three-component MCWM}
\begin{figure}[H]
	\centering
	\begin{minipage}[b]{0.45\textwidth}
		\includegraphics[width = 3in, height=2in]{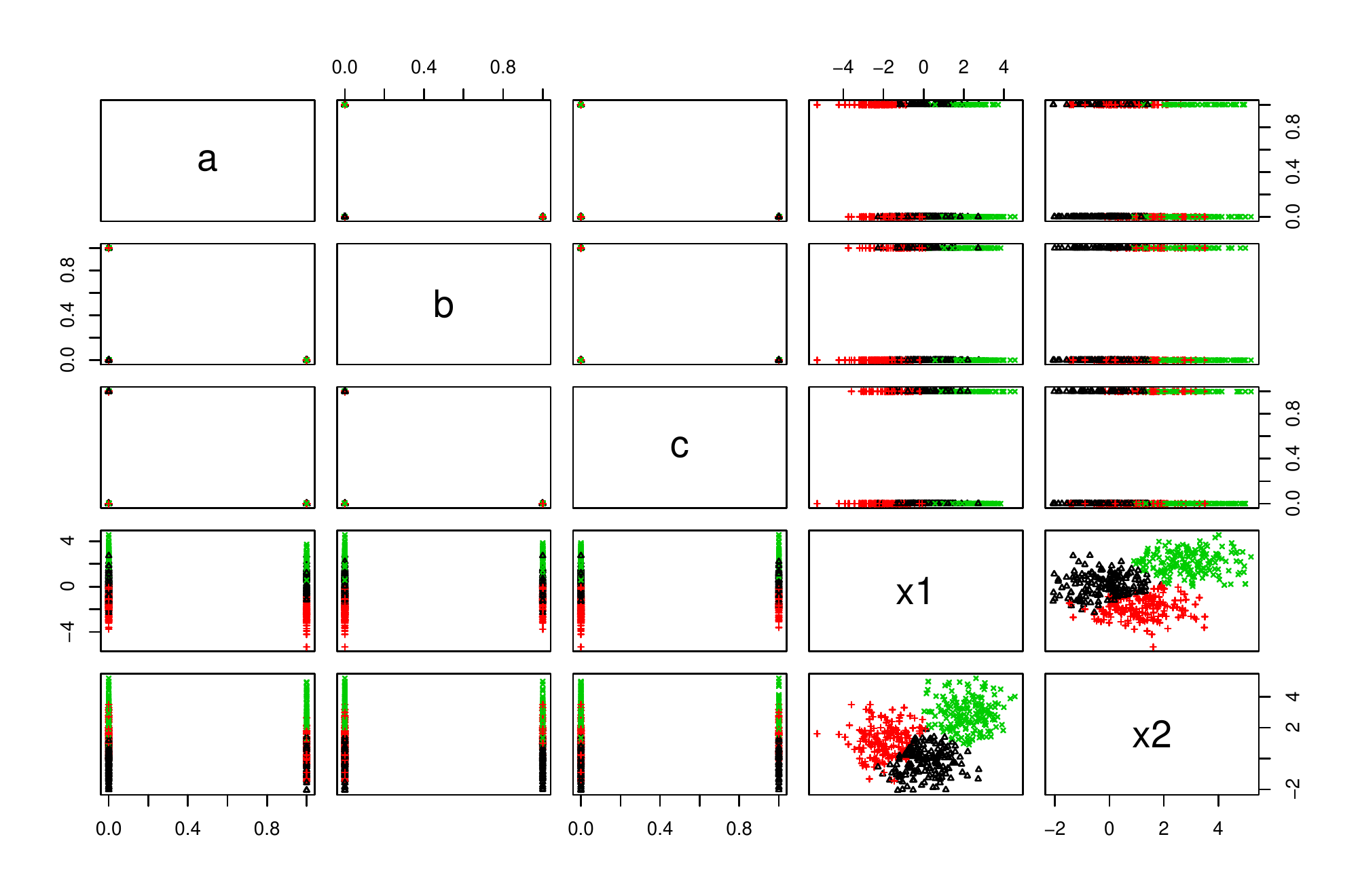}
		\centering\caption{\textit{The classification plot of multinomial CWM for ($N = 500$, $G = 3$) with two covariates.}} 
		\label{fig:7}
	\end{minipage}
	\hfil
	\begin{minipage}[b]{0.45\textwidth}
		\includegraphics[width = 3in, height=2in]{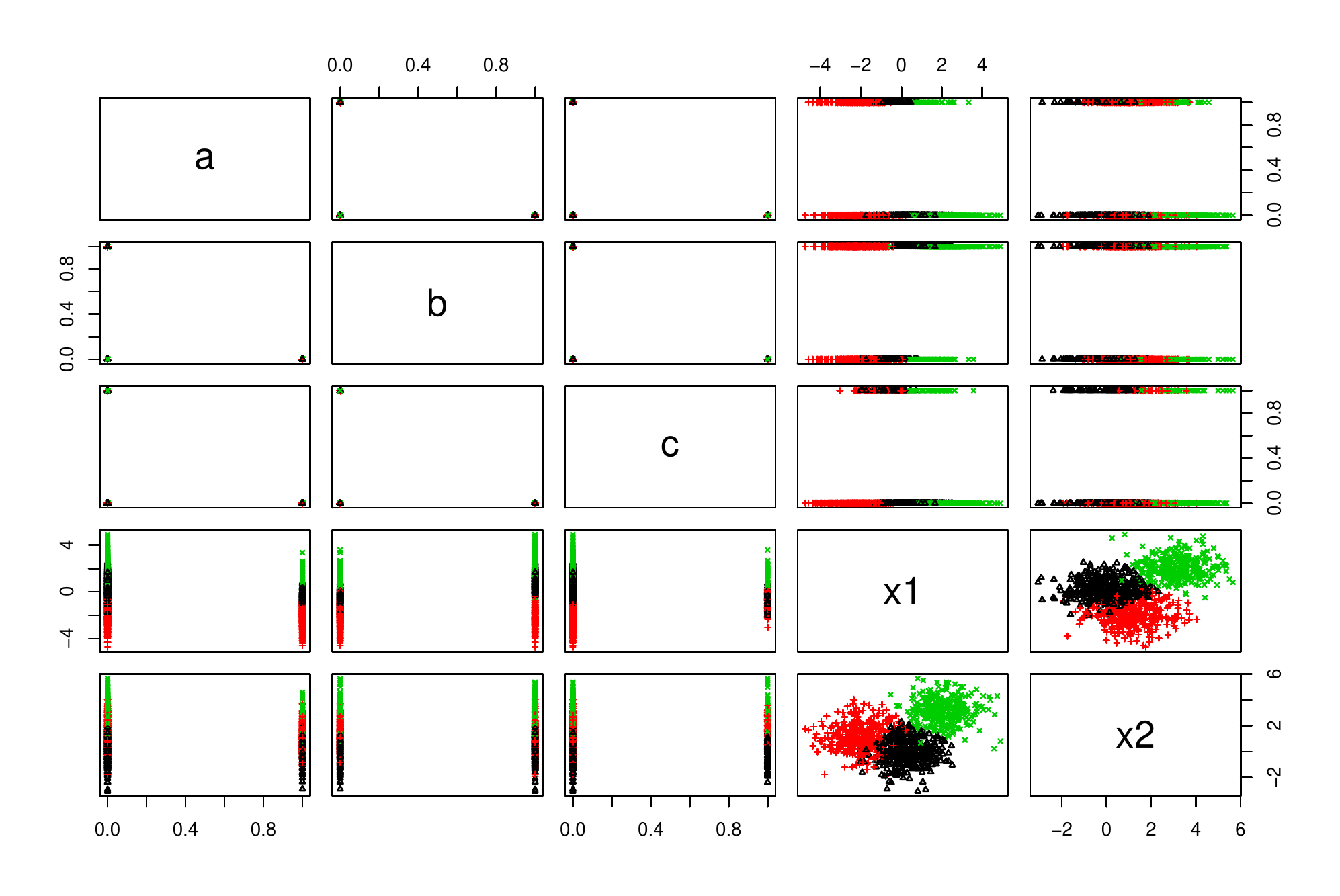}
		\caption{\textit{The classification plot of multinomial CWM for ($N = 1000$, $G = 3$) with two covariates.}} 
		\label{fig:8}
	\end{minipage}
\end{figure}

\begin{figure}[H]
	\centering
	\begin{minipage}[b]{0.45\textwidth}
		\includegraphics[width = 3in, height=2in]{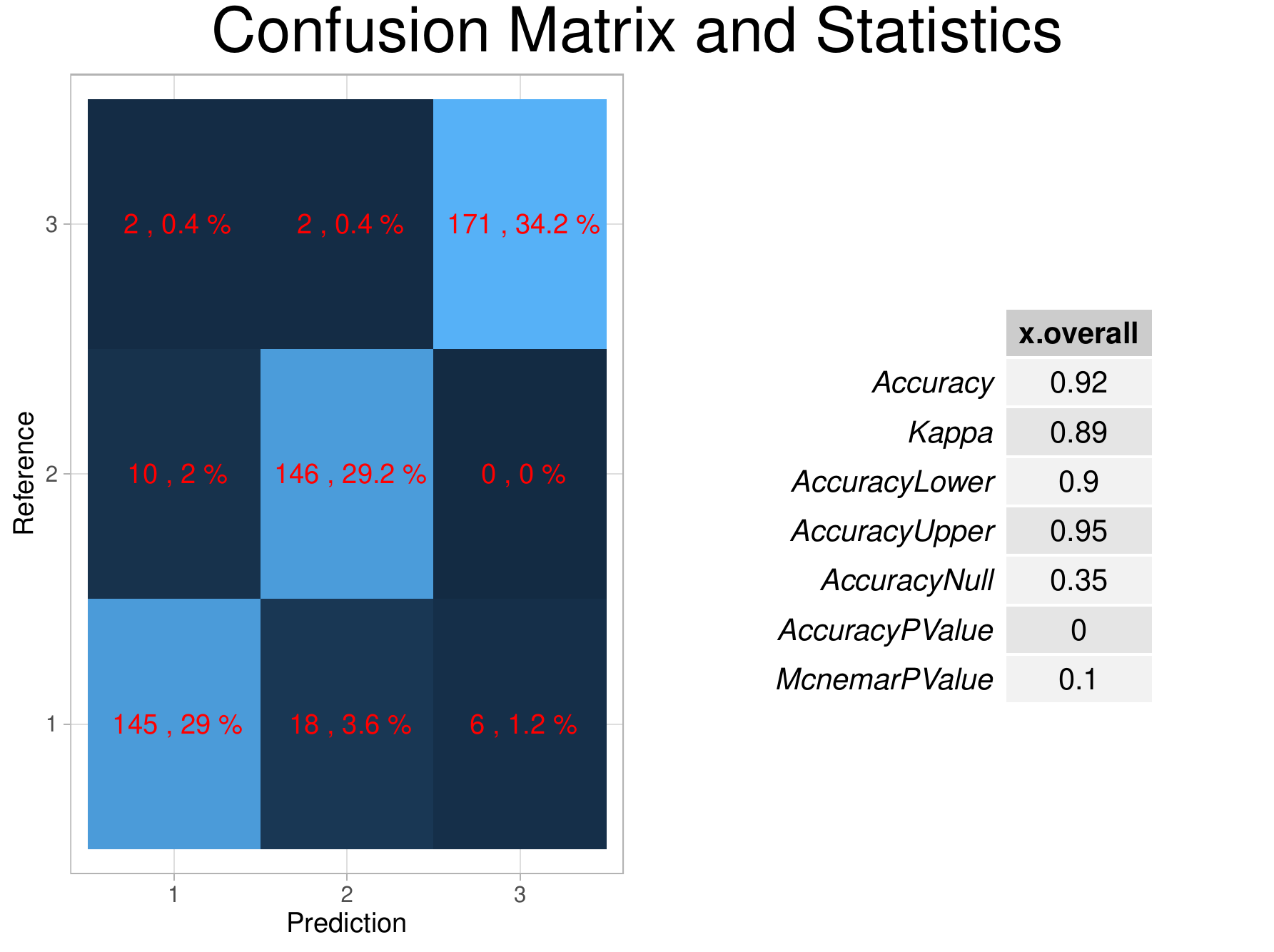}
		\centering\caption{\textit{The Visualization of Confusion Matrix and Statistics of the MCWM prediction with $N = 500$, $G = 3$.}} 
		\label{fig:9}
	\end{minipage}
	\hfil
	\begin{minipage}[b]{0.45\textwidth}
		\includegraphics[width = 3in, height=2in]{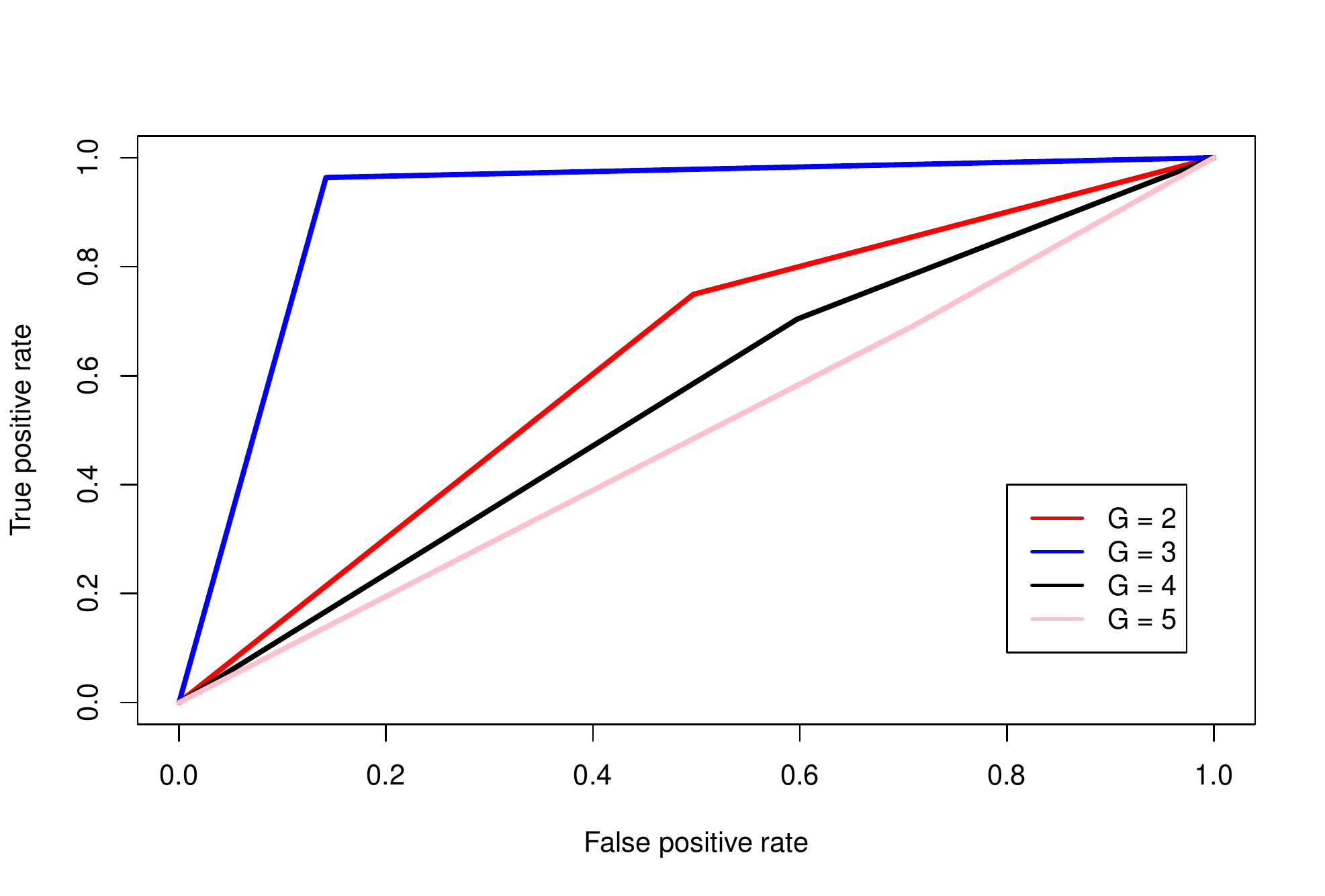}
		\caption{\textit{The Receiver's Operating Characteristic curve of the prediction by multinomialCWM ($N = 500$, $G = 3$).}} 
		\label{fig:10}
	\end{minipage}
\end{figure}

We presented the results for the three-components MCWM $G = 3$ with sample sizes $N = 500$ and $N = 1000$. Figure \eqref{fig:7} and Figure \eqref{fig:8} show the plots where each observation is clustered by the color of their classes. In Figure \eqref{fig:9}, we visualize the confusion matrix of the values presented in Table \eqref{tab:13}. Also, Figure \eqref{fig:10} shows ROC plots for sample size $N = 500$. The area under ROC curve of the three component MCWM is higher than other number of components considered in the study. In Table \eqref{tab:9}, the estimates for the mean vector, the mixing proportion, and the sigma are presented. Also, the estimates of the coefficients are shown in Table \eqref{tab:10}. Three-component MCWM provides good estimates of the parameters for both sample sizes. Table \eqref{tab:11} shows an agreement among the results of all the information criteria.
\bigskip

\begin{table}[H]
	\centering\caption{Recovered values $N = 500,1000$ and $G=3$}\vspace{0.4cm}
	\label{tab:9}
	\begin{tabular}{r@{\hspace{.2in}}r@{\hspace{.2in}}r@{\hspace{.2in}}r@{\hspace{.2in}}r@{\hspace{.2in}}r@{\hspace{.2in}}r@{\hspace{.2in}}rr}
		\hline
		&$N$&$g$&$\hat{\mu}_1$& $\hat{\mu}_2$&$\sigma^2_{11}$&$\sigma^2_{22}$&$\mathbf{\pi}$&\\ 
		\hline
		True&&$1$&$0.100$&$0.000$&$1.000$&$1.000$&$1/3$&\\ 
		&&$2$&$-2.000$&$1.000$&$1.000$&$1.000$&$1/3$&\\ 
		&&$3$&$2.000$&$3.000$&$1.000$&$1.000$&$1/3$&\\		
		\hline
		Recovered&$500$&$1$&$0.031$&$-0.179$&$0.880$&$0.870$&$0.314$&\\
		&&$2$&$-1.919$&$1.043$&$1.110$&$1.086$&$0.330$&\\
		&&$3$&$1.954$&$2.987$&$1.008$&$1.011$&$0.356$&\\
		\hline
		Recovered&$1000$&$1$&$0.235$&$0.014$&$0.901$&$1.049$&$0.340$&\\
		&&$2$&$-1.976$&$1.066$&$0.987$&$1.040$&$0.343$&\\
		&&$3$&$2.016$&$3.110$&$0.911$&$0.967$&$0.317$&\\
		\hline
	\end{tabular}
\end{table}
\begin{figure}[H]
	\centering
	\begin{minipage}[b]{0.45\textwidth}
		\includegraphics[width = 3in, height=2in]{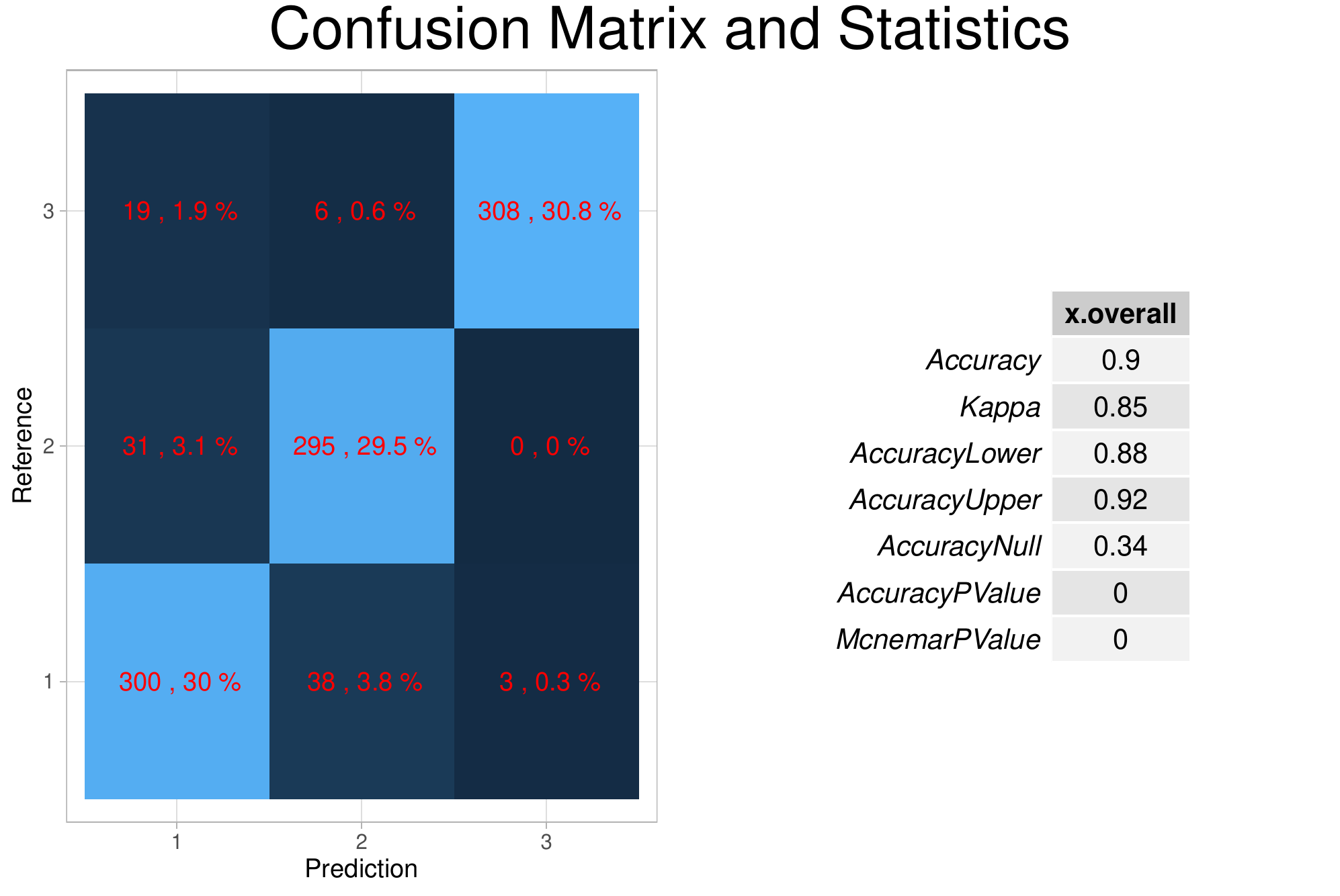}
		\centering\caption{\textit{The Visualization of Confusion Matrix and Statistics of the MCWM prediction with $N = 1000$, $G = 3$.}} 
		\label{fig:11}
	\end{minipage}
	\hfil
	\begin{minipage}[b]{0.45\textwidth}
		\includegraphics[width = 3in, height=2in]{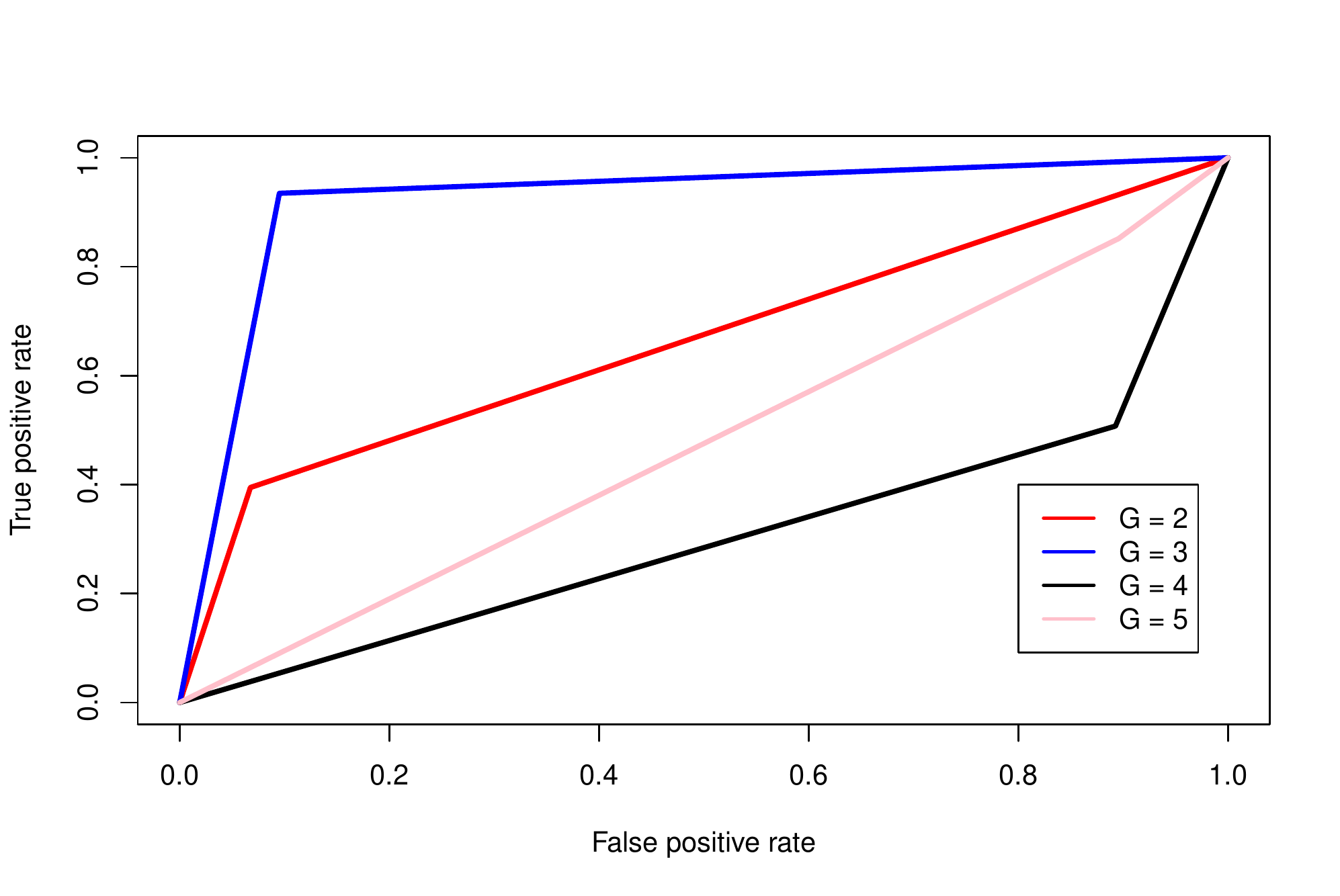}
		\caption{\textit{The Receiver's Operating Characteristic curve of the prediction by multinomialCWM ($N = 1000$, $G = 3$).}} 
		\label{fig:120}
	\end{minipage}
\end{figure}
\bigskip
\begin{table}[H]
	\centering\caption{Recovered values $N = 500,1000$ and $G=3$ with $c$ as the baseline}\vspace{0.4cm}
	\label{tab:10}
	\begin{tabular}{r@{\hspace{.2in}}r@{\hspace{.2in}}r@{\hspace{.4in}}r@{\hspace{.4in}}r@{\hspace{.3in}}r@{\hspace{.3in}}rr}
		\hline
		&$N$&$g$ &$Y$& $\hat{\beta}_0$& $\hat{\beta}_1$& $\hat{\beta}_2$&\\ 
		\hline
		True&&$1$&$a$&$0.000$&$5.000$&$0.400$&\\ 
		&&&$b$&$0.000$&$0.300$&$0.040$&\\ 
		&&$2$&$a$&$0.000$&$0.010$&$0.020$&\\
		&&&$b$&$0.000$&$2.000$&$1.000$&\\
		&&$3$&$a$&$0.000$&$1.000$&$0.030$&\\
		&&&$b$&$0.000$&$0.060$&$0.020$&\\		
		\hline
		Recovered&$500$&$1$&$a$&$0.263$&$6.236$&$0.301$&\\
		&&&$b$&$0.057$&$0.714$&$0.102$&\\
		&&$2$&$a$&$0.204$&$0.326$&$0.093$&\\
		&&&$b$&$0.232$&$2.093$&$0.651$&\\
		&&$3$&$a$&$0.051$&$0.581$&$0.103$&\\
		&&&$b$&$-0.085$&$0.417$&$-0.420$&\\
		\hline
		Recovered&$1000$&$1$&$a$&$0.135$&$4.710$&$0.385$&\\
		&&&$b$&$-0.152$&$0.044$&$-0.060$&\\ 
		&&$2$&$a$&$-0.283$&$-0.087$&$-0.039$&\\
		&&&$b$&$-0.262$&$1.426$&$0.805$&\\
		&&$3$&$a$&$-0.622$&$1.113$&$0.132$&\\
		&&&$b$&$-0.729$&$-0.134$&$0.346$&\\
		\hline
	\end{tabular}
\end{table}
\begin{table}[H]
	\centering\caption{The values of $8$ choice of Information Criteria of MCWM for different $N = 500,1000$ and $G=3$}\vspace{0.4cm}
	\label{tab:11}
	\begin{tabular}{r@{\hspace{.3in}}r@{\hspace{.1in}}r@{\hspace{.1in}}r@{\hspace{.1in}}r@{\hspace{.1in}}r@{\hspace{.1in}}r@{\hspace{.1in}}r@{\hspace{.1in}}r@{\hspace{.1in}}r@{\hspace{.1in}}r@{\hspace{.1in}}}
		\hline
		$N$&$G$&AIC&BIC&ICL&AWE&AIC3&AICc&AICu&Caic&\\ 
		\hline
		$500$&$2$&$4491.25$&$4592.40$&$4589.03$&$4072.95$&$4371.25$&$4488.73$&$4463.08$&$4270.10$&\\
		&$3$&$\mathbf{3997.76}$&$\mathbf{4149.49}$&$\mathbf{4143.77}$&$\mathbf{3370.31}$&$\mathbf{3817.76}$&$\mathbf{3992.01}$&$\mathbf{3953.57}$&$\mathbf{3666.03}$&\\
		&$4$&$4483.21$&$4685.51$&$4681.26$&$3646.61$&$4243.21$&$4471.78$&$4421.21$&$4040.91$&\\
		&$5$&$4703.66$&$4956.54$&$4951.74$&$3657.91$&$4403.66$&$4686.99$&$4621.93$&$4150.79$&\\
		\hline
		$1000$&$2$&$8459.67$&$8577.46$&$8565.33$&$8008.10$&$8339.67$&$8458.44$&$8433.12$&$8221.88$&\\
		&$3$&$\mathbf{7896.39}$&$\mathbf{8073.07}$&$\mathbf{8061.30}$&$\mathbf{7219.03}$&$\mathbf{7716.39}$&$\mathbf{7893.62}$&$\mathbf{7855.92}$&$\mathbf{7539.71}$&\\
		&$4$&$8528.43$&$8764.01$&$8752.98$&$7625.29$&$8288.43$&$8523.49$&$84.73.25$&$8052.86$&\\
		&$5$&$8737.17$&$9031.64$&$9022.30$&$7608.24$&$8437.17$&$8729.38$&$8666.44$&$8142.71$&\\
		\hline
	\end{tabular}
\end{table}

\begin{table}[H]
	\centering\caption{Adjustment Rand Index in the three component Model for $N=500$}\vspace{0.4cm}
	\label{tab:12}
	\begin{tabular}{r@{\hspace*{.3in}}r@{\hspace*{.3in}}r@{\hspace*{.3in}}r@{\hspace*{.3in}}r@{\hspace*{.3in}}r@{\hspace*{.3in}}r@{\hspace*{.3in}}rr}
		\hline
		$N$&$G$&Rand&HA&MA&FM&Jaccard&AUC&\\ 
		\hline
		$500$&$2$&$0.543$&$0.117$&$0.119$&$0.499$&$0.319$&$0.509$&\\
		&$3$&$\mathbf{0.908}$&$\mathbf{0.793}$&$\mathbf{0.794}$&$\mathbf{0.862}$&$\mathbf{0.757}$&$\mathbf{0.950}$&\\ 
		&$4$&$0.804$&$0.542$&$0.544$&$0.685$&$0.518$&$0.878$&\\
		&$5$&$0.802$&$0.530$&$0.532$&$0.672$&$0.501$&$0.709$&\\
		\hline
		$1000$&$2$&$0.614$&$0.273$&$0.273$&$0.606$&$0.410$&$0.738$&\\
		&$3$&$\mathbf{0.882}$&$\mathbf{0.734}$&$\mathbf{0.734}$&$\mathbf{0.822}$&$\mathbf{0.696}$&$\mathbf{0.930}$&\\
		&$4$&$0.860$&$0.675$&$0.675$&$0.777$&$0.633$&$0.858$&\\
		&$5$&$0.842$&$0.613$&$0.614$&$0.731$&$0.559$&$0.867$&\\
		\hline
	\end{tabular}
\end{table}
\begin{table}[H]
	\centering\caption{Confusion Matrix in the three component Model for $N=500$ and $1000$}\vspace{0.4cm}
	\label{tab:13}
	\begin{tabular}{r@{\hspace*{.3in}}r@{\hspace*{.1in}}r@{\hspace*{.1in}}r@{\hspace*{.1in}}r@{\hspace*{.1in}}rrr@{\hspace*{.1in}}r@{\hspace*{.1in}}r@{\hspace*{.1in}}r@{\hspace*{.1in}}r@{\hspace*{.1in}}r}
		\hline
		$G$&Real Component&&&$1$&&$2$&&$3$&&&Misclassification rate (\%)&\\ 
		\hline
		$500$&$1$&&&$145$&&$18$&&$6$&&&$14.20$&\\ 
		&$2$&&&$10$&&$146$&&$0$&&&$6.41$&\\
		&$3$&&&$2$&&$2$&&$171$&&&$2.29$&\\
		\hline
		rate&&&$7.60$&&&&&&&&&\\
		Accuracy&&&$92.40\%$&&&&&&&&&\\
		\hline
		$1000$&$1$&&&$300$&&$31$&&$10$&&&$14.29$&\\ 
		&$2$&&&$38$&&$295$&&$6$&&&$12.98$&\\
		&$3$&&&$3$&&$0$&&$308$&&&$0.96$&\\
		\hline
		Misclassification&&&$9.70$&\\
		Accuracy&&&$90.30\%$&\\
		\hline
	\end{tabular}
\end{table}
\bigskip
\par \noindent All the eight information criteria provide a correct selection of the number of components for both sample sizes  $500$ and $1000$. To further evaluate the model performance, we presented the values for the ARI and its variants (Table \eqref{tab:12}). Again, the model with $G = 3$ has the highest ARI and AUC values of $0.95$ and $0.93$ for both sample sizes. In Table \eqref{tab:13}, the confusion matrix was presented for $N = 500 $ at the top and $N = 1000$ at the bottom. The overall classification accuracy for $N = 500$ and $N = 1000$ are $92.4\%$ and $90.3\%$.
\begin{figure}[H]
	\centering
	\begin{minipage}[b]{0.45\textwidth}
		\includegraphics[width = 3in, height=2in]{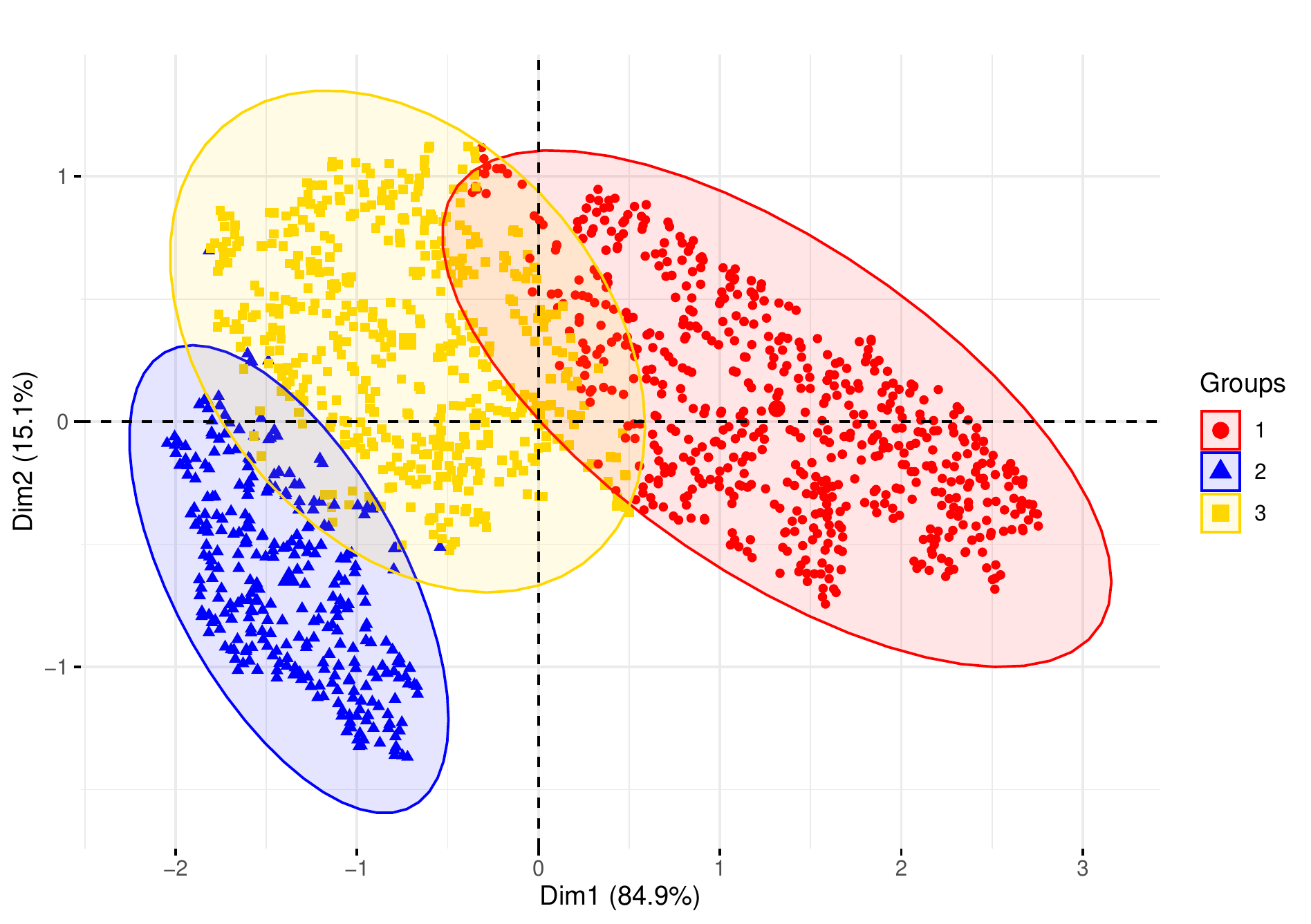}
		\centering\caption{\textit{The cluster plot of the married women using contraceptive with three levels. Cluster $1$ has $625$ married women, cluster $2$ has $316$ married women, and cluster $3$ has $511$ married women}} 
		\label{fig:13}
	\end{minipage}
	\hfil
	\begin{minipage}[b]{0.45\textwidth}
		\includegraphics[width = 3in, height=2in]{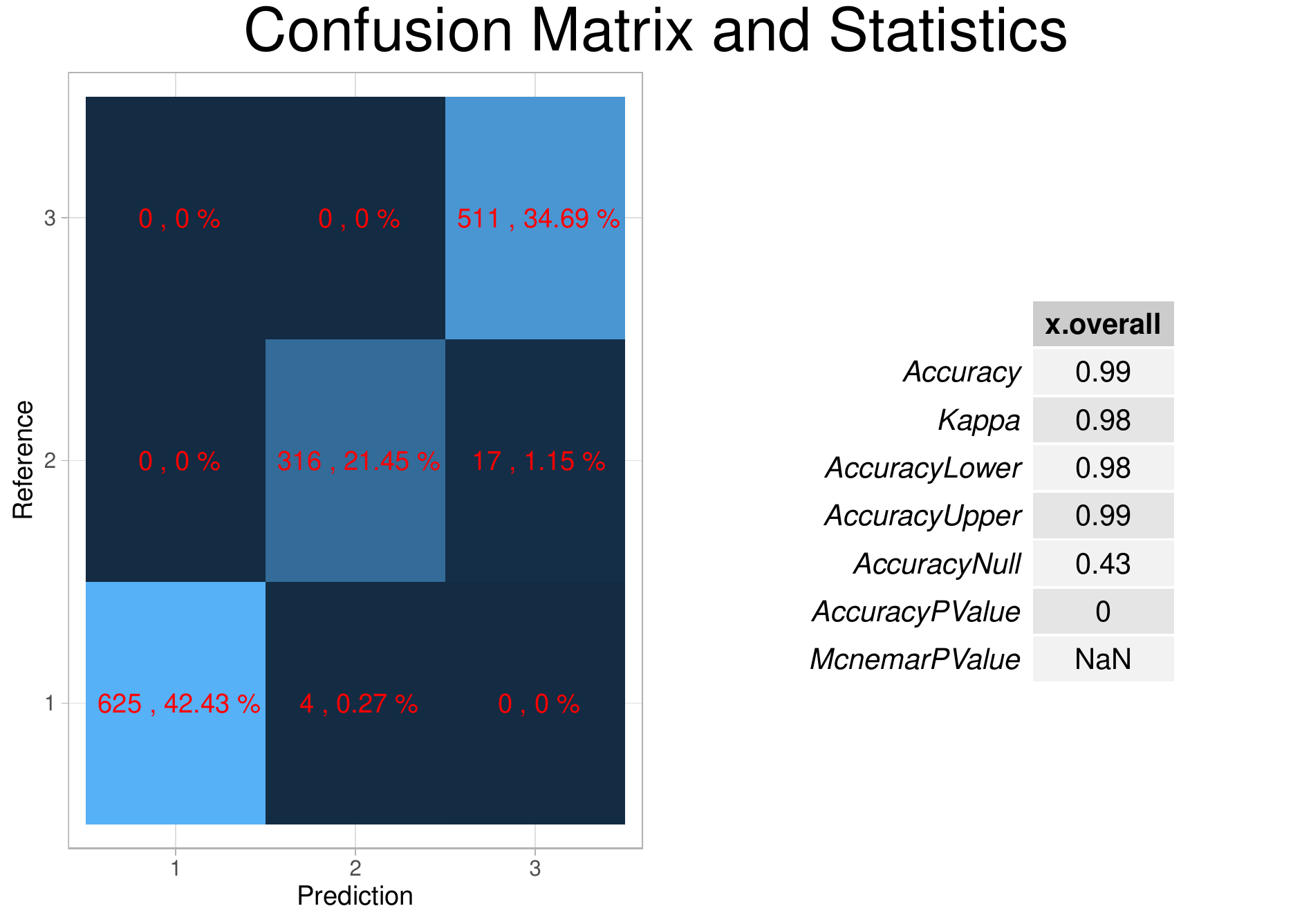}
		\centering\caption{\textit{The Visualization of Confusion Matrix and Statistics of the MCWM prediction of the use of contraceptives among married women. The result shows low confusion in the prediction of the MCWM model.}} 
		\label{fig:14}
	\end{minipage}
\end{figure}
\bigskip
\begin{table}[H]
	\centering\caption{Confusion Matrix in the three component Model for the multinomial CWM  for the use of contraceptives among married woman. MCWM has the highest prediction accuracy of about $99\%$}\vspace{0.4cm}
	\label{tab:15}
	\begin{tabular}{r@{\hspace*{.3in}}r@{\hspace*{.1in}}r@{\hspace*{.1in}}r@{\hspace*{.1in}}r@{\hspace*{.1in}}rrr@{\hspace*{.1in}}r@{\hspace*{.1in}}r@{\hspace*{.1in}}r@{\hspace*{.1in}}r}
		\hline
		Real Component&&&$1$&&$2$&&$3$&&&Misclassification rate (\%)&\\ 
		\hline
		$1$&&&$625$&&$4$&&$0$&&&$0.64$&\\ 
		$2$&&&$0$&&$316$&&$17$&&&$5.11$&\\
		$3$&&&$0$&&$0$&&$511$&&&$0.00$&\\
		\hline
		Misclassification&&&$1.43$&&&&&&&&\\
		Accuracy&&&$98.57\%$&&&&&&&&\\
		\hline
	\end{tabular}
\end{table}
\section{Real Example} \label{sec:7}
\subsubsection*{\textit{The Use of Contraceptive Among married women}}
We used the data of Contraceptive Prevalence Survey in Indonesia. The married women who were either not pregnant or do not know were included in the data \cite{Dua2017}. We aim to classify the three types of contraceptive methods (no use, long-term, or short-term) based on their characteristics.	The following information (Covariates) are as follows:
Wife's age (numerical)?, Wife's education? (categorical) 1=low, 2, 3, 4=high, Husband's education? (categorical) 1=low, 2, 3, 4=high, Number of children ever born? (numerical), Wife's religion? (binary) 0=Non-Islam, 1=Islam, Wife's now working? (binary) 0=Yes, 1=No, Husband's occupation? (categorical) 1, 2, 3, 4, Standard-of-living index? (categorical) 1=low, 2, 3, 4=high, Media exposure? (binary) 0=Good, 1=Not good, Contraceptive method used? (class attribute) 1=No-use, 2=Long-term, 3=Short-term.
\par Figure \eqref{fig:13} shows the plot of the clusters. Figure \ref{fig:14} shows the plot for the confusion matrix values. Table \eqref{tab:15} shows the cluster of the married women who used contraceptive. MCWM accurately clustered $625$ number of women in group one and $316$ women into group two. The model correctly clustered $511$ women into group three.
\begin{figure}[H]
	\centering
	\begin{minipage}[b]{0.45\textwidth}
		\includegraphics[width = 3in, height=2in]{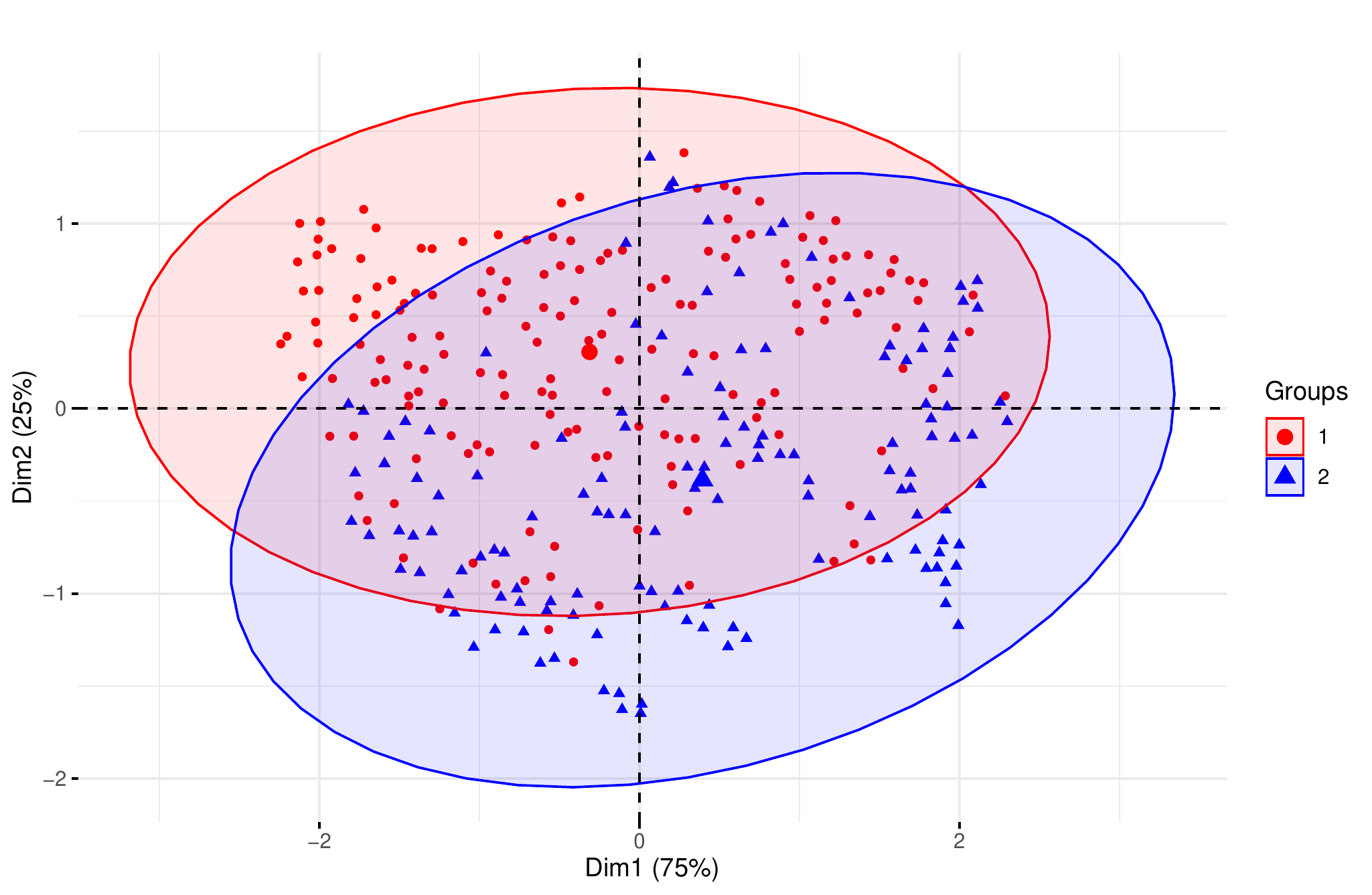}
		\centering\caption{\textit{The cluster plot of the patients diagnosed with heart disease: The class is absence (0) and presence (1,2,3,4). But according to the information criteria $G = 2$.}} 
		\label{fig:17}
	\end{minipage}
	\hfil
	\begin{minipage}[b]{0.45\textwidth}
		\includegraphics[width = 3in, height=2in]{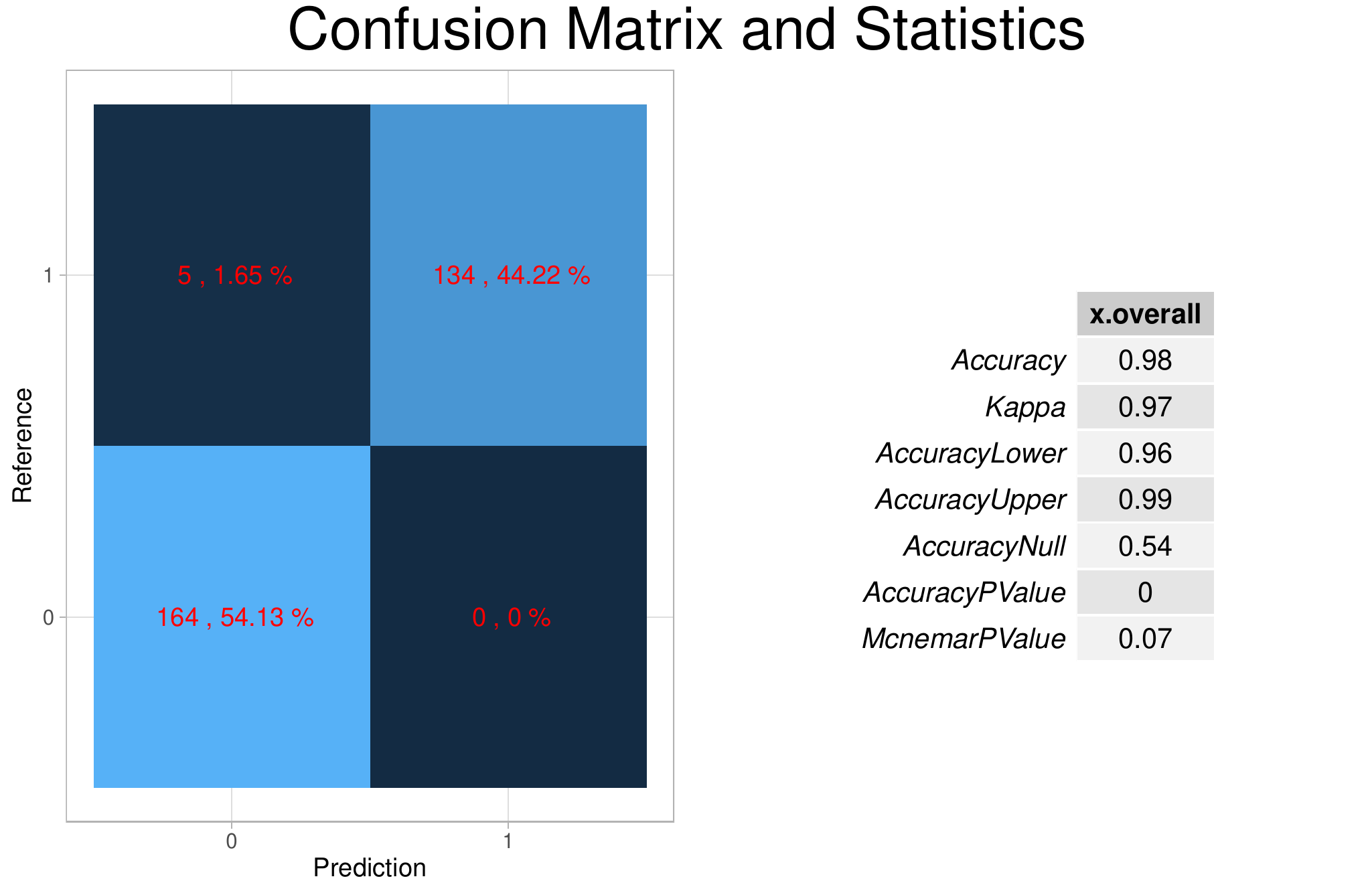}
		\caption{\textit{The Visualization of Confusion Matrix and Statistics of the MCWM prediction of the diagnosis of the patient with heart disease.}}
		\label{fig:18}
	\end{minipage}
\end{figure}
\bigskip
\begin{table}[H]
	\centering\caption{The values of eight choices of Information Criteria of MCWM for different mixture component $G$. The outcome has five levels which is absence $(0)$ and presence $(1,2,3,4)$. According to the information criteria, the number of mixture component selected is $G = 2$ which confirms what previous researchers naively suggested.}\vspace{0.4cm}
	\label{tab:14}
	\begin{tabular}{r@{\hspace{.3in}}r@{\hspace{.1in}}r@{\hspace{.1in}}r@{\hspace{.1in}}r@{\hspace{.1in}}r@{\hspace{.1in}}r@{\hspace{.1in}}r@{\hspace{.1in}}r@{\hspace{.1in}}r@{\hspace{.1in}}r@{\hspace{.1in}}}
		\hline
		&G&AIC&BIC&ICL&AWE&AIC3&AICc&AICu&Caic&\\ 
		\hline
		&$2$&$\mathbf{1358.71}$&$\mathbf{1440.41}$&$\mathbf{1435.48}$&$\mathbf{997.30}$&$\mathbf{1248.71}$&$\mathbf{1355.09}$&$\mathbf{1331.17}$&$\mathbf{1167.01}$&\\
		&$3$&$1652.25$&$1774.81$&$1773.41$&$1110.15$&$1487.25$&$1643.91$&$1607.85$&$1364.70$&\\
		&$4$&$2085.84$&$2249.24$&$2245.28$&$1363.03$&$1865.84$&$2070.49$&$2021.78$&$1702.43$&\\
		&$5$&$1982.89$&$2187.15$&$2186.65$&$1079.38$&$1707.89$&$1957.95$&$1896.04$&$1503.64$&\\
		\hline
	\end{tabular}
\end{table}
\vspace{0.05in}
\begin{table}[H]
	\centering\caption{Confusion Matrix in the three component Model for the multinomial CWM compared with other models used for the heart disease data. Among these model, MCWM has the highest prediction accuracy of about $99\%$}\vspace{0.4cm}
	\label{tab:16}
	\begin{tabular}{r@{\hspace*{.2in}}r@{\hspace*{.1in}}r@{\hspace*{.1in}}r@{\hspace*{.1in}}r@{\hspace*{.1in}}rrr@{\hspace*{.1in}}r@{\hspace*{.1in}}}
		\hline
		&Real Component&&$1$&&$2$&&Misclassification rate (\%)&\\ 
		\hline
		&$1$&&$164$&&$5$&&$7.20$&\\
		&$2$&&$0$&&$134$&&$6.61$&\\
		\hline
		Multinomial CWM&Logistic-Regression&&NTgrowth&&C4&&CLASSIT&\\
		$\vec{98.35}\%$&$77\%$&&$77\%$&&$74.8\%$&&$78.9\%$&\\
		\hline
	\end{tabular}
\end{table} 
\subsubsection*{\textit{Heart Data from Cleveland database}}
Cleveland database of Heart data contains $76$ attributes. However, many experiments suggested a subset of $14$ attributes. In particular, the Cleveland database is widely used by Machine Learning researchers (\cite{Detrano1989}, \cite{DavidandDennis1988},  and \cite{Gennari1989}). The response variable is an integer value of $0$ (absence) and $1,2,3,4$ (presence of heart diseases). Experiments with Cleveland database focused on distinguishing presence ($1,2,3,4$) from absence ($0$). The goal is to justify if the underlying group is two. We performed a model selection based on the number of mixing components in the data. Because the response variable $Y$ has five levels one might assume five groups. However, Table \eqref{tab:14} showed how all
\begin{figure}[H]
	\centering
	\includegraphics[width = 2in]{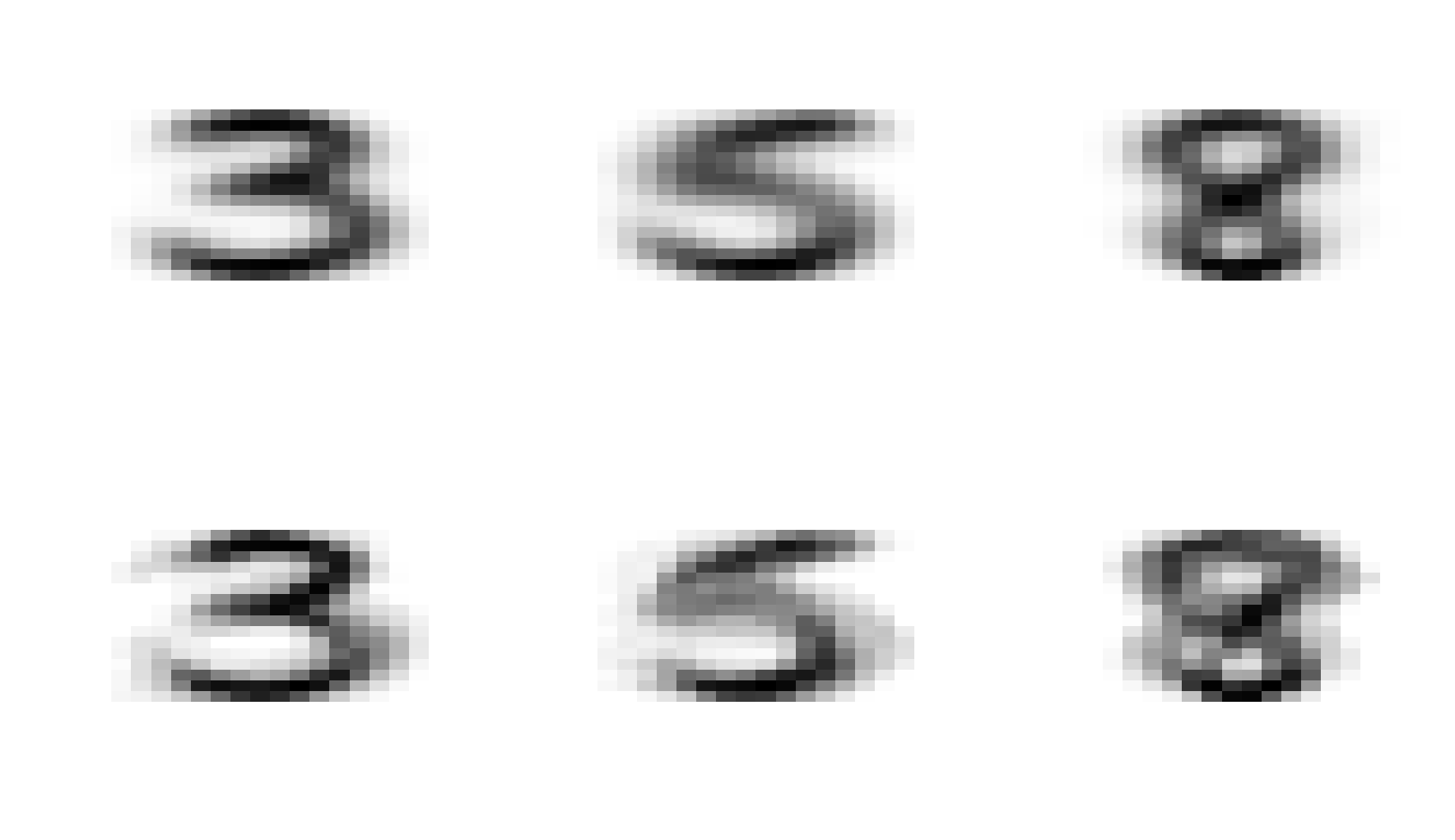}
	\caption{\textit{The original image of the data of the digits 3, 5 and 8 to be recognized is presented at the top while the means of the posterior produced by MCWM is presented at the bottom.
	}}
	\label{fig:110}
\end{figure}
\begin{table}[H]
	\centering\caption{Confusion Matrix, Accuracy, and Adjusted Rand Index of the MCWM compared with other models used for the USPS data. Among these model, MCWM has the highest prediction accuracy of about $100\%$Confusion Matrix of USPS data}\vspace{0.2cm}
	\label{Tab:6}
	\begin{tabular}{r@{\hspace*{.3in}}r@{\hspace*{.1in}}r@{\hspace*{.1in}}r@{\hspace*{.1in}}r@{\hspace*{.1in}}rr@{\hspace*{.1in}}r@{\hspace*{.1in}}r@{\hspace*{.1in}}r@{\hspace*{.1in}}rr}
		\hline
		Real Component&&&$3$&&&$5$&&$8$&\\ 
		\hline
		$3$&&&$658$&&&$0$&&$0$&&&\\ 
		$5$&&&$0$&&&$556$&&$0$&&&\\
		$8$&&&$0$&&&$0$&&$542$&&&\\
		\hline
		Accuracy&&&MCWM&&&Mclust&&HDDC&\\
		&&&$\mathbf{100.00}\%$&&&$31.89\%$&&$35.14\%$&\\
		\hline
		ARI&&&MCWM&&&Mclust&&HDDC&\\
		&&&$\mathbf{100.00}\%$&&&$63.42\%$&&$80.50\%$&\\
		\hline
	\end{tabular}
\end{table}
\bigskip
\noindent eight information criteria chose $G = 2$ rather than $G = 5$. All information criteria selected two components $G = 2$. This substantiates other works to increase the prediction accuracy of their model on the data. Table \eqref{tab:16} shows the prediction accuracy of the proposed model MCWM. We compared MCWM to previous models such as Logistic regression (\cite{Detrano1989}), NTgrowth model and C4 (\cite{DavidandDennis1988}), and CLASSIT model (\cite{Gennari1989}). Figure \eqref{fig:17} showed the cluster of patient with presence and absence of heart disease. Cluster one has $164$ patients with absence of heart disease while cluster two has $134$ patients with presence of heart disease. The actual data has $169$ patients with no heart disease and $134$ patients with heart disease. Figure \eqref{fig:18} showed that multinomial CWM has $5$ misclassifications in cluster one and achieves about $98\%$ overall. The values are presented in Table \eqref{tab:16}.
\subsection{\textit{Handwriting digit data classification}}\label{sec:1}
In this section, we compare the MCWM with model-based clustering (Mclust) (\cite{scruccaetal2016}) and High Dimensional Data Clustering (HDDC) \cite{bouveyronetal2007}. The MNIST dataset comprising of $10$-class handwritten digits, was introduced by \cite{LeCun1998}.
The full handwriting digit data is divided into training set and test set. The size of training set is $60,000 \times 784$ and the test set is $10,000\times 784$. First, we use USPS358 data set. However, due to large volume of full handwriting data set, Mclust and HDDC are not used in subsection \eqref{subsec:8.2}. This shows the limitations of the EM algorithm which is "in-scalable". 
\subsubsection*{\textit{Usps358 Data set}}
The USPS$358$ data contain only the $1,756$ images of the digits $3, 5$, and $8$. These digits are the most difficult to discriminate. Each digit is a $16\times 16$ gray level image and is represented as a $256$-dimensional vector in the USPS358 data set. Table ~\eqref{Tab:6} showed confusion matrix, Accuracy, and Adjusted Rand Index. We showed how MCMC outperforms Mclust and HDDC. The Accuracy of MCWM is $100\%$ while Mclust achieves $31.89\%$ and HDDC achieves $35.14\%$. This is also confirmed by the result of the Adjusted Rand Index that counts
\begin{figure}[H]
	\centering
	\includegraphics[width = 3in]{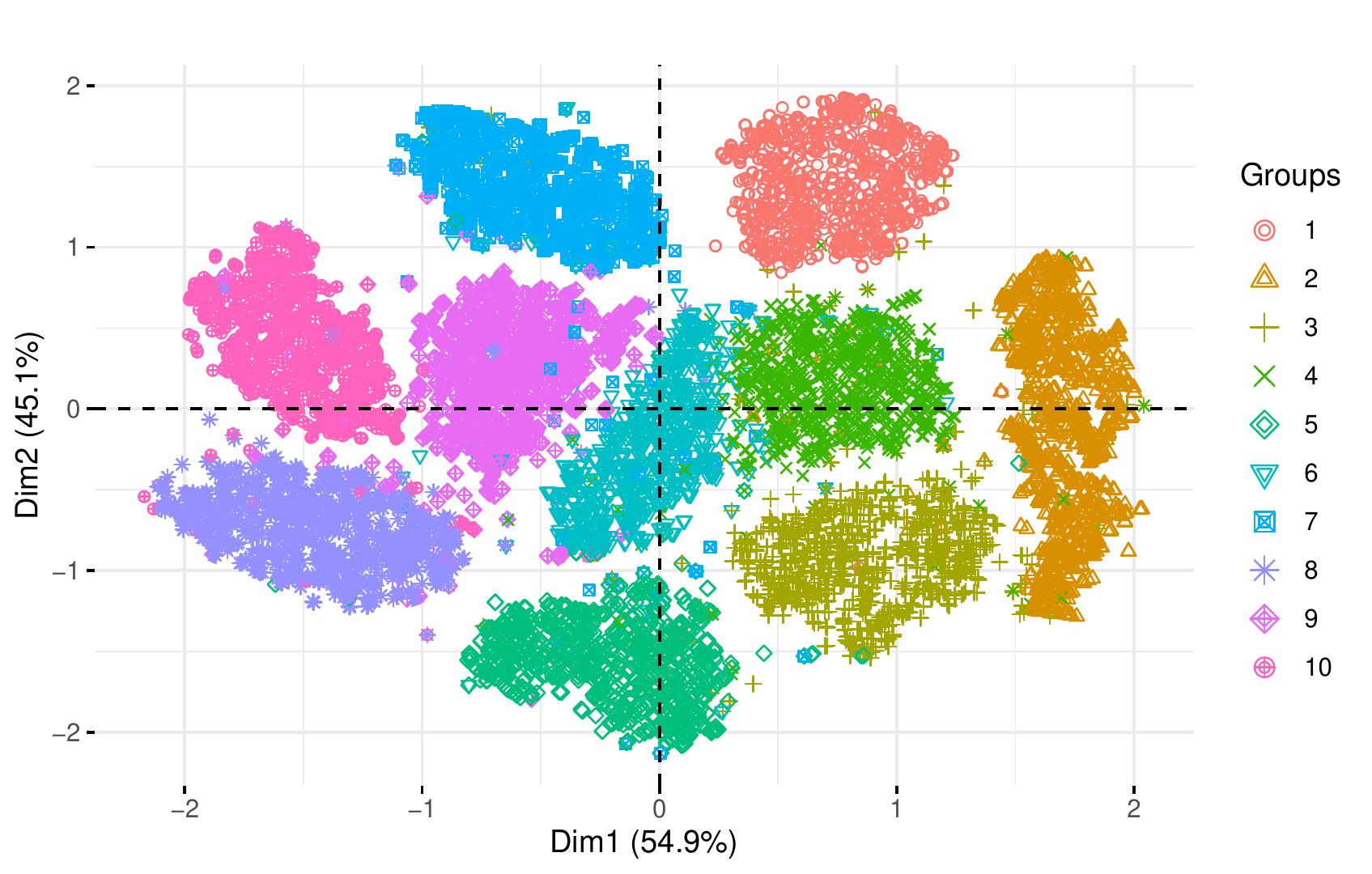}
	\caption{The MCWM classification plot for Handwritten image data set}
	\label{fig:12}
\end{figure}
\bigskip
\begin{figure}[H]
	\centering
	\includegraphics[width = 0.8in]{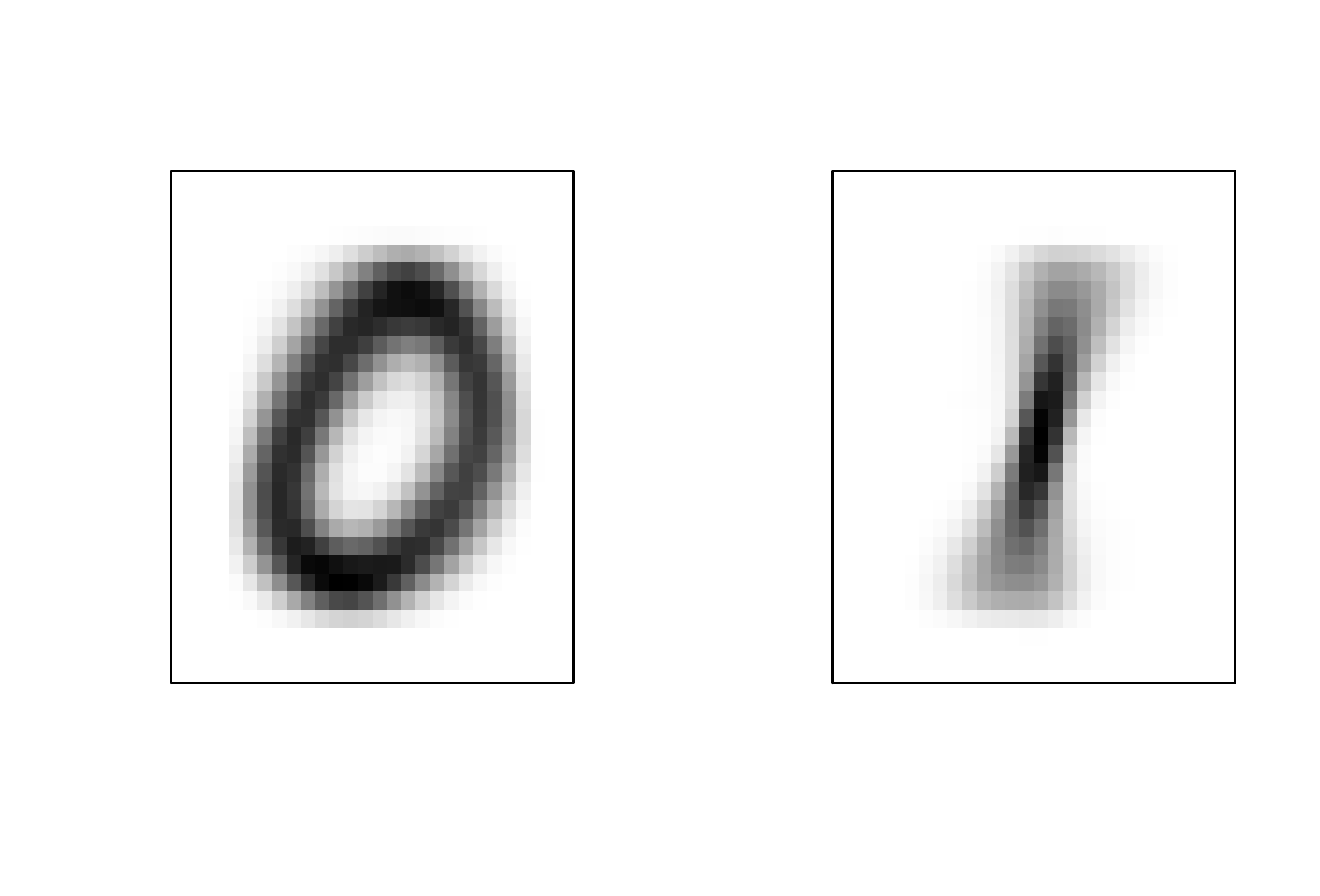}
	\includegraphics[width = 0.8in]{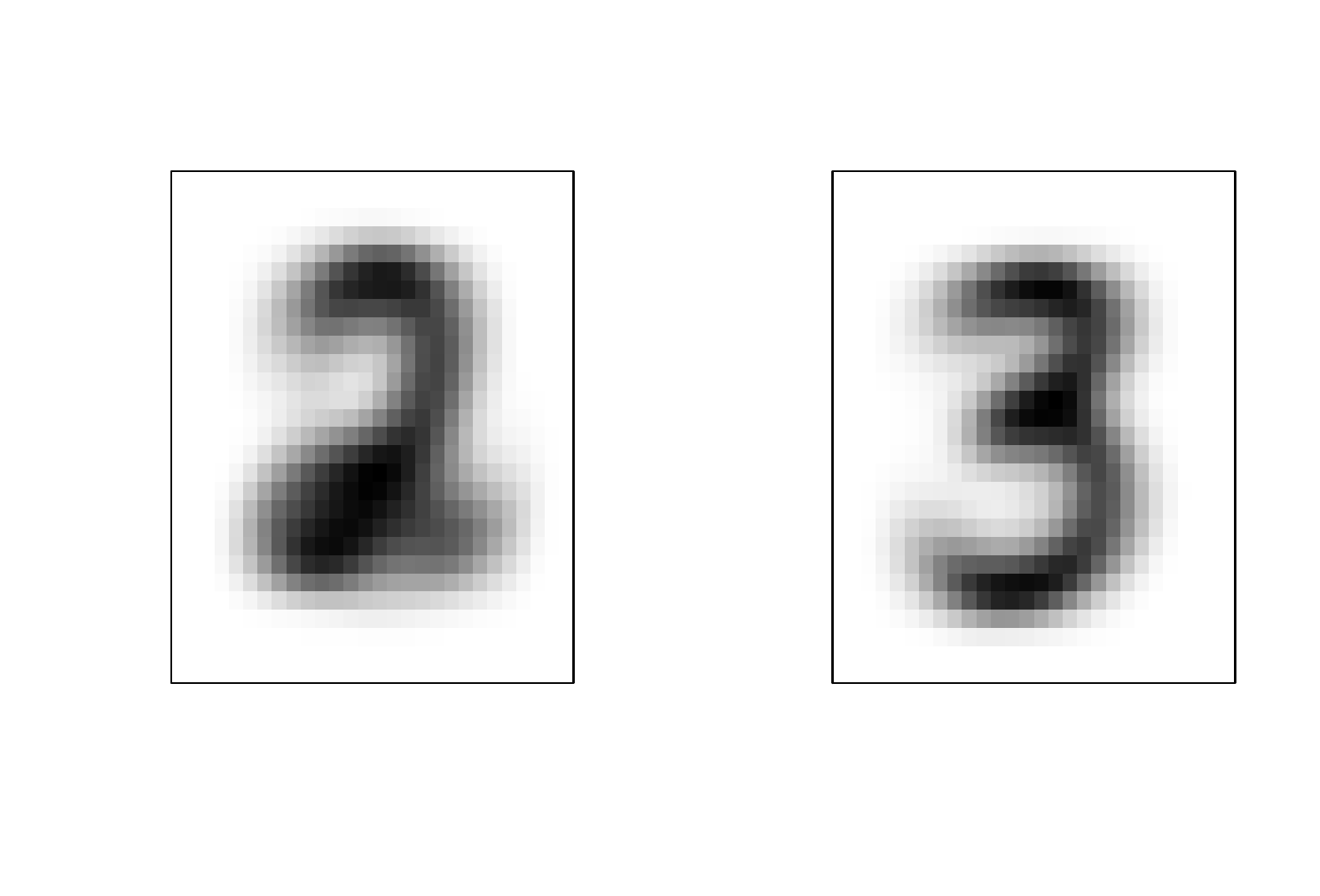}
	\includegraphics[width = 0.8in]{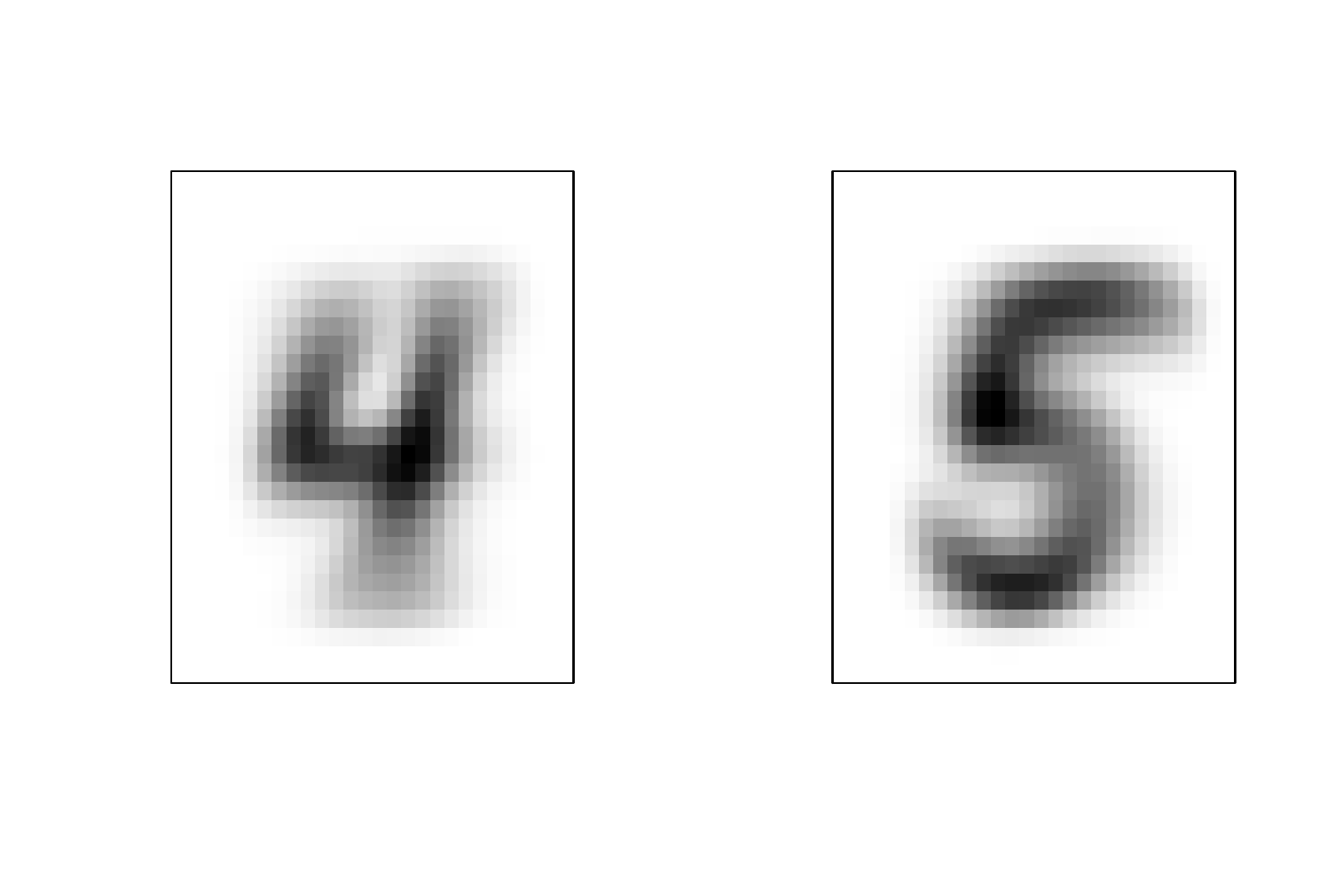}
	\includegraphics[width = 0.8in]{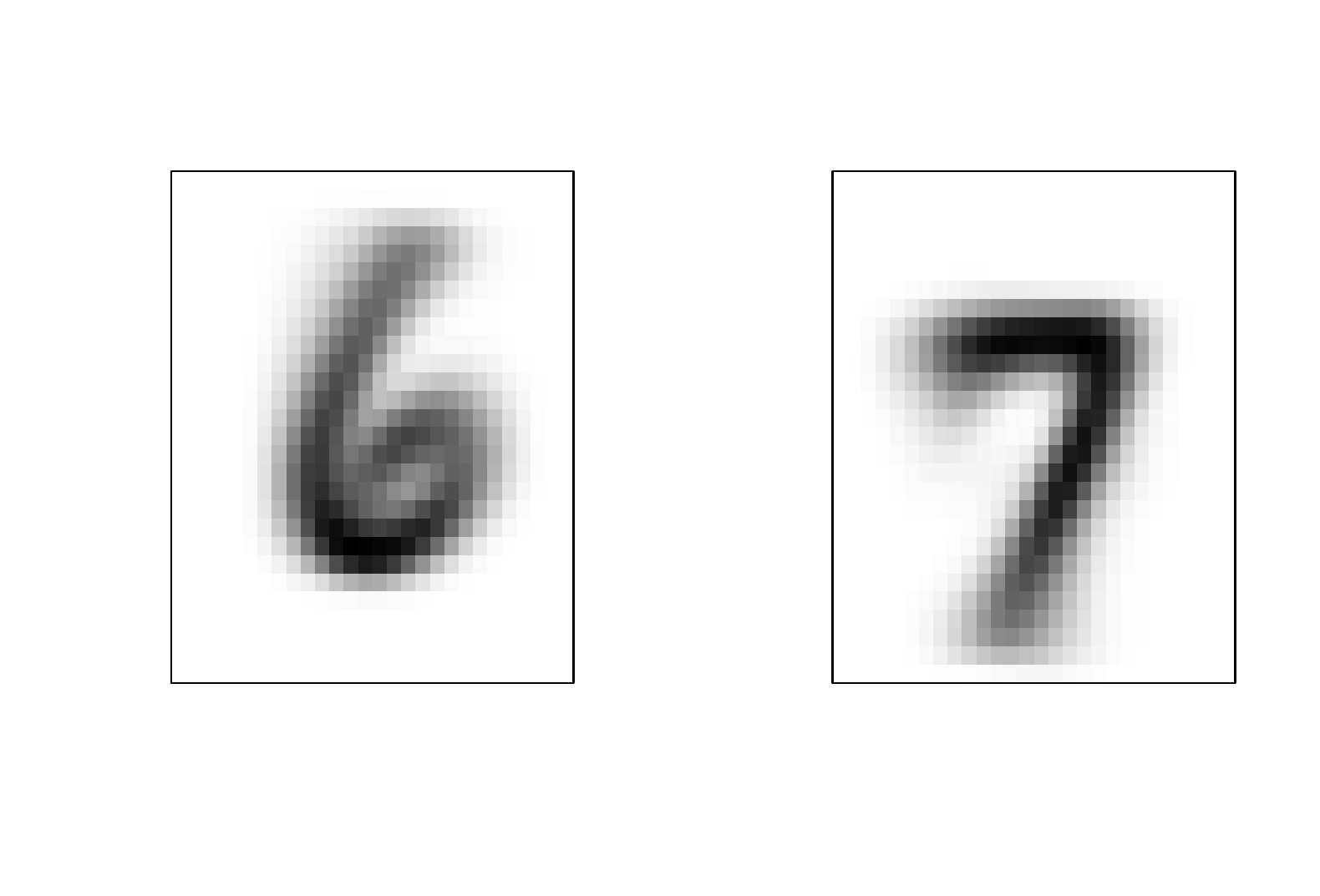}
	\includegraphics[width = 0.8in]{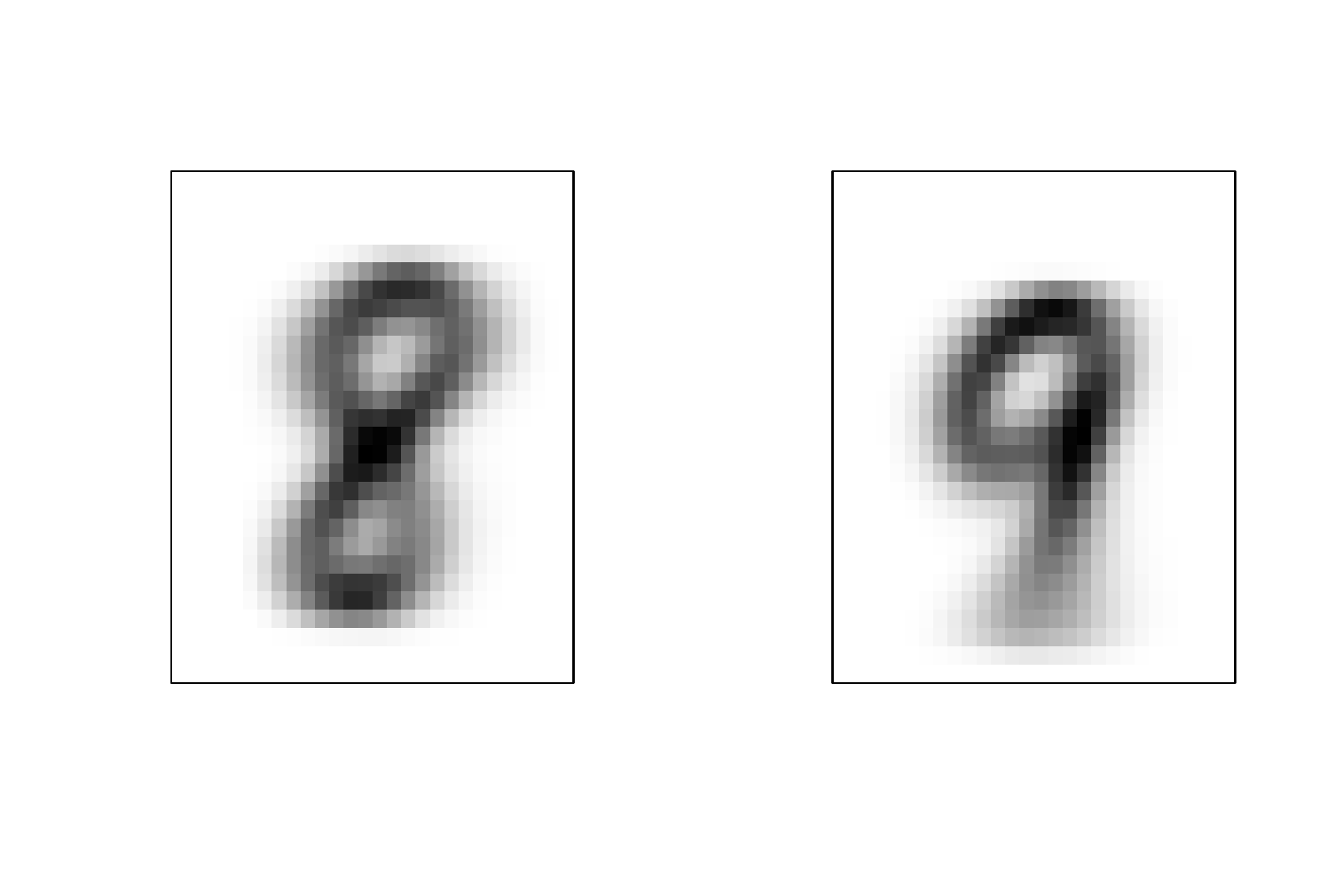}
	\caption{Image recognized from 10 groups}
	\label{fig:112}
\end{figure}
\bigskip
\noindent the number of correctly classification between the actual and the predicted. Figure ~\eqref{fig:110} (top) showed a sample of handwriting digits from the USPS postal services (USPS358 data set in the MBCbook R package). And Figure ~\eqref{fig:110} (below) are the images of the different handwriting.
\subsubsection*{\textit{Full Handwriting Image classification}}\label{subsec:8.2}
The results provided in Table ~\eqref{Tab:8} is the confusion matrix for both training set and test set. In Table ~\eqref{Tab:8}, the accuracy provided by MCWM is about $93\%$ while it achieved the prediction accuracy of about $92\%$. We note here that no comparison was made with both Mclust and HDDC due to the volume of the data. The use of batch size technique makes MCWM scalable to high dimensional data. Also, we computed the Adjusted Rand Index for the training set and test set to be $81\%$ and $80\%$ respectively.
Figure ~\eqref{fig:112} showed a well partitioned classes of the numbers which is the output of MCWM.
\section{Conclusion and Future Work}\label{sec:8}
In this paper, we introduced a new cluster weighted models called a Multinomial CWM for clustering data with multiclass response variables. We used the softmax function as the probability of the multinomial distribution. MCWM extended preexisting Binomial CWM for binary response variables \cite{Ingrassiaetal2015}. Different from the previous work in the field of cluster weighted models, MCWM allows modeling of multinomial response variables. Furthermore, we described through a simulation study an EM algorithm via Iteratively Re-weighted Least Squares for parameter estimation. To investigate the performance of the model, several performance metrics were used such as confusion matrix, Receiver's Operating Characteristics (ROC), and ARI with its variants. We also showed through different eight information criteria that the MCWM is able to discover the mixture components hidden in the data. Following the conditions of identifiability, we can conclude that MCWM is identifiable. We provided different supporting plots such as the classification plots, confusion matrix plots, and ROC plot. We also evaluated the performance of the proposed model on two real datasets. We showed that Multinomial CWM has higher accuracy compared to other models for the data. However, from the perspective of the EM via IRLS algorithm for the parameter
\begin{table}[H]
	\centering\caption{Confusion Matrix of the MCWM compared with other models used for the training set (Top) and test set (Bottom) of Handwritten Image}\vspace{0.2cm}
	\label{Tab:8}
	\begin{tabular}{rrrrrrrrrrrrr}
		\hline
		&class 1 & class 2 & class 3 & class 4 & class 5 & class 6 & class 7 & class 8 & class 9 & class 10 &\\ 
		\hline
		class 1 & 5807 &   1 &  14 &  20 &   8 &  19 &   8 &   8 &  28 &  10 &\\ 
		class 2 &   1 & 6629 &  19 &  22 &   2 &  13 &   1 &  17 &  23 &  15 &\\ 
		class 3 &  25 &  82 & 5298 & 178 &  61 &  28 &  45 &  74 & 137 &  30 &\\ 
		class 4 &  19 &  19 &  67 & 5673 &   6 & 178 &   5 &  46 &  61 &  57 &\\ 
		class 5 &  26 &  26 &  26 &  10 & 5383 &  10 &  26 &  31 &  25 & 279 &\\ 
		class 6 &  53 &  25 &  34 & 236 &  29 & 4820 &  49 &  18 & 120 &  37 &\\ 
		class 7 &  73 &  26 &  69 &  16 &  52 &  99 & 5549 &   8 &  19 &   7 &\\ 
		class 8 &  17 &  16 &  42 &  26 &  31 &  10 &   1 & 5965 &   6 & 151 &\\ 
		class 9 &  24 & 200 &  45 & 328 &  16 & 273 &  28 &  29 & 4784 & 124 &\\ 
		class 10 &  25 &  20 &  17 &  54 &  87 &  42 &   2 & 194 &  16 & 5492 &\\ 
		\hline
		& class 1 & class 2 & class 3 & class 4 & class 5 & class 6 & class 7 & class 8 & class 9 & class 10 & \\ 
		\hline
		class 1 & 964 &   0 &   1 &   2 &   1 &   4 &   3 &   3 &   1 &   1 &\\ 
		class 2 &   0 & 1120 &   2 &   4 &   0 &   1 &   3 &   2 &   3 &   0 &\\ 
		class 3 &   8 &  19 & 898 &  35 &  10 &   3 &  11 &  14 &  31 &   3 &\\ 
		class 4 &   4 &   0 &   6 & 942 &   1 &  28 &   1 &  12 &   8 &   8 &\\ 
		class 5 &   1 &   2 &   6 &   4 & 897 &   0 &   5 &  10 &  10 &  47 &\\ 
		class 6 &  11 &   4 &   3 &  51 &   6 & 769 &   9 &   7 &  25 &   7 &\\ 
		class 7 &  17 &   3 &  10 &   6 &  11 &  30 & 880 &   1 &   0 &   0 &\\ 
		class 8 &   2 &   7 &  16 &   9 &   3 &   1 &   0 & 957 &   2 &  31 &\\ 
		class 9 &   9 &  22 &   5 &  46 &   6 &  64 &   8 &  16 & 780 &  18 &\\ 
		class 10 &  10 &   8 &   0 &  10 &  15 &  10 &   0 &  31 &   1 & 924 &\\ 
		\hline
	\end{tabular}
\end{table}
\bigskip
\noindent  estimation of MCWM, the main limitations encountered are in-scalability to large dataset due to slow-to-convergence nature of EM algorithm and the problem of initialization. The matrix inversion from the multinomial distribution estimation computed by the iteratively re-weighted least squares often led to the problem of singularity in covariance matrix. Singularity at the M-step often caused the EM algorithm to explode. Additionally, the use of conventional maximum likelihood techniques at the M-step become so unstable due to covariance matrix inversion. We optimized the M-step with Stochastic Gradient Descent algorithm (EM-SGD) which atoned for the "Curse of high-dimensionality". In the future, we wish to explore different variants of EM algorithm for this type of proposed model such as EM algorithm via Simulated Annealing (SA) (EM-SA). EM-SA algorithm will avoid the problem of local maxima in EM algorithm. With the help of EM-SGD, we also wish in the future to work in the perspective of Deep Learning where MCWM will be computed at each hidden unit.
\subsection*{Appendix A}
\subsection*{Proof of Identifiability for MCMW}
The proof is divided into two parts. The first part is built upon results given in \cite{Hennig2000} while the second part is built upon the results given in \cite{GrunandLeisch2008}. Consider the class of models defined in Equation \eqref{equ:3} and prove the equality as follows;
\begin{equation}
	\mathlarger\sum_{g=1}^{G} F(\mathbf{y}|\mathbf{x};\mathbf{\beta}_j^g)\phi_d(\mathbf{x};\mathbf{\mu}_g, \mathbf{\Sigma}_g)\pi_g = \mathlarger\sum_{k=1}^{\tilde{G}} F(\mathbf{y}|\mathbf{x};\mathbf{\tilde{\beta}}_j^k)\phi_d(\mathbf{x};\mathbf{\tilde{\mu}}_k, \mathbf{\tilde{\Sigma}}_k)\tilde{\pi}_k
	\label{equ:19}
\end{equation}
holds for almost all $\mathbf{x} \in \mathcal{R}^d$ and for all $\mathbf{y} \in \mathcal{Y}$ iff $G = \tilde{G}$ and there exists a one-to-one correspondence between $\{1,...,G\}$ and $\{1,...,\tilde{G}\}$ such that for each $g \in \{1,...,G\}$ there exists a correspondent element $k \in \{1,..., \tilde{G}\}$ and $\mathbf{\beta}^g_j = \mathbf{\tilde{\beta}}^k_j$, $\mathbf{\mu}_g = \mathbf{\tilde{\mu}}_k$, $\mathbf{\Sigma}_g = \mathbf{\tilde{\Sigma}}_k$, and $\pi_g = \tilde{\pi}_k$. Integrating both sides of Equation \eqref{equ:13} over $\mathcal{Y}$ as follows
\begin{equation}
	\mathlarger \int_{\mathcal{Y}} \hspace{0.05in}\mathlarger\sum_{g=1}^{G} F(\mathbf{y}|\mathbf{x};\mathbf{\beta}_j^g)\phi_d(\mathbf{x};\mathbf{\mu}_g, \mathbf{\Sigma}_g)\pi_g \hspace{0.05in} d\mathbf{y}= \mathlarger \int_{\mathcal{Y}}\hspace{0.05in} \mathlarger\sum_{k=1}^{\tilde{G}} F(\mathbf{y}|\mathbf{x};\mathbf{\tilde{\beta}}_j^k)\phi_d(\mathbf{x};\mathbf{\tilde{\mu}}_k, \mathbf{\tilde{\Sigma}}_k)\tilde{\pi}_k \hspace{0.05in} d\mathbf{y}
	\label{equ:20}
\end{equation} 
\begin{equation}
	\mathlarger\sum_{g=1}^{G}\phi_d(\mathbf{x};\mathbf{\mu}_g, \mathbf{\Sigma}_g)\pi_g\hspace{0.05in}\mathlarger \int_{\mathcal{Y}} F(\mathbf{y}|\mathbf{x};\mathbf{\beta}_j^g) \hspace{0.05in} d\mathbf{y}=\mathlarger\sum_{k=1}^{\tilde{G}} \phi_d(\mathbf{x};\mathbf{\tilde{\mu}}_k, \mathbf{\tilde{\Sigma}}_k)\tilde{\pi}_k\hspace{0.05in}\mathlarger\int_{\mathcal{Y}} F(\mathbf{y}|\mathbf{x};\mathbf{\tilde{\beta}}_j^k) \hspace{0.05in} d\mathbf{y}
	\label{equ:21}
\end{equation} 
Since 
\begin{equation}
	\int_{\mathcal{Y}} F(\mathbf{y}|\mathbf{x};\mathbf{\beta}_j^g) \hspace{0.05in} d\mathbf{y} = \mathlarger\int_{\mathcal{Y}} F(\mathbf{y}|\mathbf{x};\mathbf{\tilde{\beta}}_j^k) \hspace{0.05in} d\mathbf{y} = 1 \notag
\end{equation}
then equation \ref{equ:21} becomes
\begin{equation}
	\mathlarger\sum_{g=1}^{G}\phi_d(\mathbf{x};\mathbf{\mu}_g, \mathbf{\Sigma}_g)\pi_g = \mathlarger\sum_{k=1}^{\tilde{G}} \phi_d(\mathbf{x};\mathbf{\tilde{\mu}}_k, \mathbf{\tilde{\Sigma}}_k)\tilde{\pi}_k 
	\label{equ:22}
\end{equation} 
Let us set 
$p(\mathbf{x};\mathbf{\tilde{\mu}}, \mathbf{\tilde{\Sigma}},\tilde{\pi}) = \mathlarger\sum_{k=1}^{\tilde{G}} \phi_d(\mathbf{x};\mathbf{\tilde{\mu}}_k, \mathbf{\tilde{\Sigma}}_k)\tilde{\pi}_k$ and $p(\mathbf{x};\mathbf{\mu}, \mathbf{\Sigma},\pi)=\mathlarger\sum_{k=1}^{G} \phi_d(\mathbf{x};\mathbf{\mu}_k, \mathbf{\Sigma}_k)\pi_k$, where $(\mathbf{\mu, \Sigma}, \mathbf{\pi}) = \{(\mathbf{\mu}_g, \mathbf{\Sigma}_g, \pi_g); g = 1,...,G\}$, $(\tilde{\mathbf{\mu}}, \tilde{\mathbf{\Sigma}}, \tilde{\mathbf{\pi}}) = \{(\tilde{\mathbf{\mu}}_k, \tilde{\mathbf{\Sigma}}_k, \tilde{\pi}_k); k = 1,...,\tilde{G}\}$. Applying the Bayes' theorem gives
$p(\mathcal{D}_g|\mathbf{x};\tilde{\mathbf{\mu}},\tilde{\mathbf{\Sigma}}, \tilde{\mathbf{\pi}}) = \frac{\phi_d(\mathbf{x};\tilde{\mathbf{\mu}}_k,\tilde{\mathbf{\Sigma}}_k)\tilde{\pi}_k}{\mathlarger\sum_{t=1}^{\tilde{G}}\phi_d(\mathbf{x};\tilde{\mathbf{\mu}}_t,\tilde{\mathbf{\Sigma}}_t)\tilde{\pi}_t}$ and $p(\mathcal{D}_g|\mathbf{x};\mathbf{\mu},\mathbf{\Sigma}, \mathbf{\pi}) = \frac{\phi_d(\mathbf{x};\mathbf{\mu}_g,\mathbf{\Sigma}_g)\pi_g}{\mathlarger\sum_{s=1}^{G}\phi_d(\mathbf{x};\mathbf{\mu}_s,\mathbf{\Sigma}_s)\pi_s}$. Then Substituting Equation \eqref{equ:17} and Equation \eqref{equ:18} into Equation \eqref{equ:19} and Equation \eqref{equ:20} we get $p(\mathcal{D}_k|\mathbf{x};\tilde{\mathbf{\mu}},\tilde{\mathbf{\Sigma}}, \tilde{\mathbf{\pi}}) = \frac{\phi_d(\mathbf{x};\tilde{\mathbf{\mu}}_k,\tilde{\mathbf{\Sigma}}_k)\tilde{\pi}_k}{p(\mathbf{x};\tilde{\mathbf{\mu}}, \tilde{\mathbf{\Sigma}},\tilde{\mathbf{\pi}})}$ and 
$p(\mathcal{D}_g|\mathbf{x};\mathbf{\mu},\mathbf{\Sigma}, \mathbf{\pi}) = \frac{\phi_d(\mathbf{x};\mathbf{\mu}_g,\mathbf{\Sigma}_g)\pi_g}{p(\mathbf{x};\mathbf{\mu}, \mathbf{\Sigma},\mathbf{\pi})}$
thus, Equation \eqref{equ:3} can be written as 
\begin{equation}
	p(\mathbf{x},\mathbf{y};\mathbf{\Theta}) = p(\mathbf{x};\mathbf{\mu},\mathbf{\Sigma},\mathbf{\pi})\mathlarger\sum_{g=1}^{G} F(\mathbf{y};\mathbf{M},\mathbf{\theta}^g) p(\mathcal{D}_g|\mathbf{x};\mathbf{\mu},\mathbf{\Sigma}, \mathbf{\pi}) = p(\mathbf{x};\mathbf{\mu},\mathbf{\Sigma},\mathbf{\pi})p(\mathbf{y}|\mathbf{x};\mathbf{\varphi})
	\label{equ:23}
\end{equation}
where
\begin{equation}
	p(\mathbf{y}|\mathbf{x};\mathbf{\varphi}) = \mathlarger\sum_{g=1}^{G} F(\mathbf{y};\mathbf{M},\mathbf{\theta}^g)p(\mathcal{D}_g|\mathbf{x};\mathbf{\mu},\mathbf{\Sigma}, \mathbf{\pi})
	\label{equ:24}
\end{equation}
where also the positive weights $\gamma_g(\mathbf{x})$ in Equation \eqref{equ:16} can be written as $\gamma_g(\mathbf{x}) = p(\mathcal{D}_g|\mathbf{x};\mathbf{\mu},\mathbf{\Sigma}, \mathbf{\pi})$. 
To complete the first part of the proof, since the $p(\mathcal{D}_g|\mathbf{x};\mathbf{\mu}, \mathbf{\Sigma}, \mathbf{\pi})$ and $p(\mathcal{D}_k|\mathbf{x};\tilde{\mathbf{\mu}},\tilde{\mathbf{\Sigma}}, \tilde{\mathbf{\pi}})$ are defined according to the Equation \eqref{equ:21} and Equation \eqref{equ:22}, we get:
\begin{equation}
	\pi_g = \mathlarger \int_{\mathcal{X}} \phi_d(\mathbf{x};\vec{\mu}_g, \mathbf{\Sigma}_g)\pi_g \hspace{0.05in}d\mathbf{x} = \mathlarger \int_{\mathcal{X}} \frac{\phi_d(\mathbf{x};\mathbf{\mu}_g, \mathbf{\Sigma}_g)\pi_g}{\mathlarger \sum_{s=1}^{G}\phi_d(\mathbf{x};\mathbf{\mu}_s, \mathbf{\Sigma}_s)\pi_s}\hspace*{0.05in}\mathlarger \sum_{s=1}^{G}\phi_d(\mathbf{x};\mathbf{\mu}_s, \mathbf{\Sigma}_s)\pi_s\hspace{0.05in}d\mathbf{x}\notag
\end{equation}
since $p(\mathcal{D}_g|\mathbf{x};\mathbf{\mu},\mathbf{\Sigma}, \mathbf{\pi}) = p(\mathcal{D}_k|\mathbf{x};\tilde{\mathbf{\mu}},\tilde{\mathbf{\Sigma}}, \tilde{\mathbf{\pi}})$, then
\begin{equation}
	= \mathlarger \int_{\mathcal{X}} \frac{\phi_d(\mathbf{x};\tilde{\mathbf{\mu}}_k,\tilde{\mathbf{\Sigma}}_k)\tilde{\pi}_k}{\mathlarger\sum_{t=1}^{\tilde{G}}\phi_d(\mathbf{x};\tilde{\mathbf{\mu}}_t,\tilde{\mathbf{\Sigma}}_t)\tilde{\pi}_t}\hspace*{0.05in}\mathlarger \sum_{t=1}^{\tilde{G}}\phi_d(\mathbf{x};\tilde{\mathbf{\mu}}_t, \tilde{\mathbf{\Sigma}}_t)\tilde{\pi}_t\hspace{0.05in}d\mathbf{x}
	= \tilde{\pi}_k\mathlarger \int_{\mathcal{X}} \phi_d(\mathbf{x};\tilde{\mathbf{\mu}}_k,\tilde{\mathbf{\Sigma}}_k)\hspace{0.05in}d\mathbf{x} 
	\label{equ:25}
\end{equation}
since $\int_{\mathcal{X}} \phi_d(\mathbf{x};\tilde{\mathbf{\mu}}_k,\tilde{\mathbf{\Sigma}}_k)\hspace{0.05in}d\mathbf{x} = 1$, then $\pi_g = \tilde{\pi}_k$.
Following the same step, it can be seen that $\mathbf{\mu}_g = \tilde{\mathbf{\mu}}_k$ and $\mathbf{\Sigma}_g = \tilde{\mathbf{\Sigma}}_k$.
\par The class of models in Equation \eqref{equ:24} is identifiable if the condition of intra-component label switching is fulfilled, this builds up the second part of the proof.
\begin{equation}
	\mathlarger\sum_{g=1}^{G} p(\mathcal{D}_g|\mathbf{x};\mathbf{\mu},\mathbf{\Sigma}, \mathbf{\pi}) \mathlarger \prod_{i} \mathlarger\prod_{j} F(\mathbf{y}_{ij};M_{ij},\mathbf{\theta}_{ij}^g) = 	\mathlarger\sum_{k=1}^{\tilde{G}} p(\mathcal{D}_g|\mathbf{x};\tilde{\mathbf{\mu}},\tilde{\mathbf{\Sigma}}, \tilde{\mathbf{\pi}}) \mathlarger \prod_{i} \mathlarger\prod_{j} F(\mathbf{y}_{ij};M_{ij},\tilde{\mathbf{\theta}}_{ij}^k)
	\label{equ:26}
\end{equation}
implies that $G = \tilde{G}$ and there exists a one-to-one mapping between the two sets $\{1,...,G\}$ and $\{1,...,\tilde{G}\}$ such that $\mathbf{\theta}^g = \tilde{\mathbf{\theta}}^k$. Moreover the relationship between response variable $\mathbf{y}$ and the covariates $\mathbf{x}$ is $\mathbf{\theta}$.
\begin{equation}
	\ln\bigg(\frac{\mathbf{\theta}_j}{\mathbf{\theta}_J}\bigg) = \mathbf{x}'_i\mathbf{\beta}_j
	\label{equ:27}
\end{equation}
Exponentiating Equation \eqref{equ:27} gives $e^{\mathbf{x}'_i\mathbf{\beta}_j}$ and since $\mathbf{\theta}_j = \tilde{\mathbf{\theta}}_j$ then $\mathbf{\theta}_J = \tilde{\mathbf{\theta}}_J$ which implies that 
$e^{\mathbf{x}'_i\mathbf{\beta}_j} = e^{\mathbf{x}'_i\tilde{\mathbf{\beta}}_j}$. Now following \cite{GrunandLeisch2008}, we show that $e^{\mathbf{x}'_i(\mathbf{\beta}_j - \tilde{\mathbf{\beta}}_j)} = c_j$. Following \cite{Ingrassiaetal2015}, we introduce two sets
\begin{equation}
	\mathcal{X} = \bigg\lbrace\mathbf{x}\in \mathcal{R}^d:\hspace{0.05in}\text{for each} \hspace{0.05in} g,w \in \{1,...,G\} \hspace{0.05in} \text{and} \hspace{0.05in} k,h \in \{1,...,\tilde{G}\}: e^{\mathbf{x}'_i\mathbf{\beta}^g_j} = e^{\mathbf{x}'_i\mathbf{\beta}^w_j} \implies \mathbf{\theta}^g_j =  \mathbf{\theta}^w_j, \notag
	\notag
\end{equation}
\begin{equation}
	\mathbf{\beta}^g_j = \mathbf{\beta}^w_j, e^{\mathbf{x}'_i\mathbf{\beta}^g_j} = e^{\mathbf{x}'_i\tilde{\mathbf{\beta}}^k_j} \implies \mathbf{\theta}^g_j =  \tilde{\mathbf{\theta}}^k_j, \mathbf{\beta}^g_j = \tilde{\mathbf{\beta}}^k_j, e^{\mathbf{x}'_i\tilde{\mathbf{\beta}}^k_j} = e^{\mathbf{x}'_i\tilde{\mathbf{\beta}}^h_j} \implies \tilde{\mathbf{\theta}}^k_j =  \tilde{\mathbf{\theta}}^h_j, \tilde{\mathbf{\beta}}^k_j = \tilde{\mathbf{\beta}}^h_j\bigg\rbrace
	\label{equ:28}
\end{equation}
Since $\mathbf{\theta}^g \neq \mathbf{\theta}^w$ for $g \neq w$. The following holds for all $i \in \mathbf{I}_k$ and for $j = 1,..., J-1$, $y_{ij} = \delta_{ij}$ and $y_{iJ} = M_{ij} - y_{ij}$, then Kronecker delta $\delta_{ij} = 1$ if $i = j$ and zero otherwise. The multinomial coefficients on both side of equation \ref{equ:24} are canceled
\begin{equation}
	e^{\mathbf{x}'_i(\mathbf{\beta}^g-\mathbf{\beta}^k)} = \frac{\mathlarger \sum_{g=1}^{G} p(\mathcal{D}_g|\mathbf{x};\mathbf{\mu},\mathbf{\Sigma},\mathbf{\pi})\bigg[e^{\mathbf{x}'_i\mathbf{\beta}^g}\mathlarger \prod_{i}\mathlarger\prod_{j}\bigg(\mathlarger\sum_{u=1}^{J}e^{\vec{x}'_i\mathbf{\beta}^g_u}\bigg)^{-M_{ij}}\bigg]}{\mathlarger \sum_{k=1}^{\tilde{G}} p(\mathcal{D}_k|\mathbf{x};\vec{\mu},\mathbf{\Sigma},\mathbf{\pi})\bigg[e^{\mathbf{x}'_i\tilde{\mathbf{\beta}}^k}\mathlarger \prod_{i}\mathlarger\prod_{j}\bigg(\mathlarger\sum_{u=1}^{J}e^{\vec{x}'_i\tilde{\mathbf{\beta}}^k_u}\bigg)^{-M_{ij}}\bigg]}
	\label{equ:30}
\end{equation}
for a fixed $\mathbf{x} \in \mathcal{X}$, according to equation \ref{equ:22}, $\gamma_1(\mathbf{x}), ..., \gamma_G(\mathbf{x})$ which is also $p(\mathcal{D}_1|\mathbf{x};\mathbf{\mu},\mathbf{\Sigma},\mathbf{\pi}), .., p(\mathcal{D}_G|\mathbf{x};\mathbf{\mu},\mathbf{\Sigma},\mathbf{\pi})$ and  $p(\mathcal{D}_1|\mathbf{x};\tilde{\mathbf{\mu}},\tilde{\mathbf{\Sigma}},\tilde{\mathbf{\pi}}), .., p(\mathcal{D}_{\tilde{G}}|\mathbf{x};\tilde{\mathbf{\mu}},\tilde{\mathbf{\Sigma}},\tilde{\mathbf{\pi}})$ which all sum to one. It follows that, for each $\mathbf{x} \in \mathcal{X}$, the density given in Equation \eqref{equ:24} is then identifiable if and only $G = \tilde{G}$ and there exists $k \in \{1,...,\tilde{G}\}$ such that $\mathbf{\theta}^g = \tilde{\mathbf{\theta}}^k$ and $p(\mathcal{D}_G|\mathbf{x};\mathbf{\mu},\mathbf{\Sigma},\mathbf{\pi}) = p(\mathcal{D}_{\tilde{G}}|\mathbf{x};\tilde{\mathbf{\mu}},\tilde{\mathbf{\Sigma}},\tilde{\mathbf{\pi}})$, then the left hand side Equation \eqref{equ:30} is constant for all $i \in \mathbf{I}_g$.
\subsection*{Appendix B}
\subsection*{Derivation of Maximization via IRLS}\label{app3} 
The updated estimates $\psi^{(q+1)}$ are the solutions of the following M-step.
\begin{equation}
	\frac{\partial}{\partial \mathbf{\beta}}Q_1(\mathbf{\Omega};\mathbf{\psi}^{(q)}) = \frac{\partial}{\partial \mathbf{\beta}}\mathlarger\sum_{i=1}^{N}\mathlarger\sum_{g=1}^{G}z^{(q)}_{ig} \Bigg\lbrace y_{i1}\ln\phi_{i1g} + \mathlarger\sum_{j=2}^{J}y_{ij}\ln\phi_{ijg}\Bigg\rbrace
	\label{equ:B.1}
\end{equation}
Following from Equation (\eqref{equ:7}) and $f(\mathbf{x},\mathbf{\beta}_{jg}) = \mathbf{\beta}_{0jg} + \mathbf{x}'\mathbf{\beta}_{1jg}$, the update of $\mathbf{\psi}^{(q)}$ of Equation (\eqref{equ:B.1}) is derived as follows
\begin{equation}
	\mathbf{\beta}^{(q+1)}_{jg} = \mathbf{\beta}^{(q)}_{jg} + [I(\mathbf{\beta}^{(q)}_{jg})]^{-1}S(\mathbf{\beta}^{(q)}_{jg})
	\label{equ:B.2}
\end{equation}
where $I(\mathbf{\beta}^{(q)}_{jg})$ is the Fisher information matrix and $S(\mathbf{\beta}^{(q)}_{jg})$ is the score function. The parameters are estimated as follows:
\begin{equation}
	\frac{\partial}{\partial \mathbf{\beta}}Q_1(\mathbf{\Omega};\mathbf{\psi}^{(q)}) = \frac{\partial}{\partial \mathbf{\phi}}Q_1(\mathbf{\Omega};\mathbf{\psi}^{(q)})\frac{\partial}{\partial f(\mathbf{x};\mathbf{\beta})}\mathbf{\phi}(f)\frac{\partial}{\partial \mathbf{\beta}}f(\mathbf{x};\mathbf{\beta})
	\label{equ:B.3}
\end{equation}
Maximizing Equation (\eqref{equ:B.1}) with respect to $\mathbf{\beta}$ is equivalent to independently maximizing each $J$ class and $G$ component expressions
\begin{equation}
	S(\mathbf{\beta}^{(q)}_{jg}) =  \mathlarger\sum_{i=1}^{N} z^{(q)}_{ig} \Bigg\lbrace\frac{y_{ij}}{\phi_{ijg}} -\frac{y_{i1}}{\phi_{i1g}} \Bigg\rbrace
	\label{equ:B.4}
\end{equation} 
using the expression $y_{i1} = n_i - y_{ij}$ and $\phi_{i1g} = 1 - \phi_{ijg}$ Equation (\eqref{equ:B.4}) can be written as 
\begin{equation}
	S(\mathbf{\beta}^{(q)}_{jg}) = \frac{\partial}{\partial \phi_{ijg}}Q_1(\mathbf{\Omega};\mathbf{\psi}^{(q)}) = \mathlarger\sum_{i=1}^{N}z^{(q)}_{ig} \Bigg\lbrace\frac{y_{ij}}{\phi_{ijg}} -\frac{n_i - y_{ij}}{1-\phi_{ijg}} \Bigg\rbrace \frac{\partial}{\partial\mathbf{\beta}_{jg}}\mathbf{\phi}_{ijg}
	\label{equ:B.5}
\end{equation} 
The next equation will be derived on element-by-element basis, that is;
\begin{equation}
	\frac{\partial}{\partial f_{ijg}(\mathbf{x};\mathbf{\beta})}\mathbf{\phi}(f_{ijg}) = \frac{\partial}{\partial f_{ijg}(\mathbf{x};\mathbf{\beta})} \frac{\exp\{f_{ijg}(\mathbf{x};\mathbf{\beta}_{jg})\}}{\sum_j^J \exp\{f_{ijg}(\mathbf{x};\mathbf{\beta}_{jg})\}}
	\label{equ:B.6}
\end{equation}
\begin{equation}
	= \frac{\exp\Big\{f_{ijg}\big(\mathbf{x};\mathbf{\beta}_{jg}\big)\Big\} \mathlarger \sum_{j = 1}^{J} \exp\Big\{f_{ijg}\big(\mathbf{x};\mathbf{\beta}_{jg}\big)\Big\} - \exp\Big\{f_{ijg}\big(\mathbf{x};\mathbf{\beta}_{jg}\big)\Big\} \exp\Big\{f_{ijg}\big(\mathbf{x};\mathbf{\beta}_{jg}\big)\Big\}}{\Bigg(\mathlarger \sum_{j = 1}^{J} \exp\Big\{f_{ijg}\big(\mathbf{x};\mathbf{\beta}_{jg}\big)\Big\}\Bigg)^2}
	\label{equ:B.7}
\end{equation}
then we have,
\begin{equation}
	= \frac{\exp\Big\{f_{ijg}\big(\mathbf{x};\mathbf{\beta}_{jg}\big)\Big\}}{\mathlarger \sum_{j = 1}^{J} \exp\Big\{f_{ijg}\big(\mathbf{x};\mathbf{\beta}_{jg}\big)\Big\}}  \Bigg(1 - \frac{\exp\Big\{f_{ijg}\big(\mathbf{x};\mathbf{\beta}_{jg}\big)\Big\}}{\mathlarger \sum_{j = 1}^{J} \exp\Big\{f_{ijg}\big(\mathbf{x};\mathbf{\beta}_{jg}\big)\Big\}}\Bigg)
	\label{equ:B.8}
\end{equation}
Equation (\eqref{equ:B.6}) is $\phi_{ijg}(1-\phi_{ijg})\vec{x}_i$.
The score function becomes 
\begin{equation}
	S(\mathbf{\beta}^{(q)}_{jg}) =  \mathlarger\sum_{i=1}^{N} z^{(q)}_{ig}\mathbf{x}_i \phi^{(q)}_{ijg}(1-\phi^{(q)}_{ijg})\Bigg\lbrace\frac{y_{ij}}{\phi^{(q)}_{ijg}} -\frac{y_{i1}}{\phi^{(q)}_{i1g}} \Bigg\rbrace.
	\label{equ:B.9}
\end{equation} 
Now, we derive the Fisher information matrix as follows;
\begin{equation}
	S(\mathbf{\beta}^{(q)}_{jg}) =  \mathlarger\sum_{i=1}^{N} z^{(q)}_{ig}\mathbf{x}_i \phi^{(q)}_{ijg}\phi^{(q)}_{i1g} \Bigg\lbrace\frac{y_{ij}}{\phi^{(q)}_{ijg}} -\frac{y_{i1}}{\phi^{(q)}_{i1g}} \Bigg\rbrace
	\label{equ:B.10}
\end{equation} 
\begin{equation}
	S(\mathbf{\beta}^{(q)}_{jg}) =  \mathlarger\sum_{i=1}^{N} z^{(q)}_{ig}\mathbf{x}_i\bigg\lbrace y_{ij}\phi^{(q)}_{i1g} - y_{i1} \phi^{(q)}_{ijg} \bigg\rbrace
	\label{equ:B.11}
\end{equation} 
using the expression $y_{i1} = n_i - y_{ij}$ and $\phi_{i1g} = 1 - \phi_{ijg}$ again Equation (\eqref{equ:B.11}) becomes
\begin{equation}
	=  \mathlarger\sum_{i=1}^{N} z^{(q)}_{ig}\mathbf{x}_i\bigg\lbrace y_{ij} - n_i \phi^{(q)}_{ijg} \bigg\rbrace
	\label{equ:B.12}
\end{equation} 
\begin{equation}
	-\frac{\partial}{\partial\beta_{jg}} S(\mathbf{\beta}^{(q)}_{jg}) =  \mathlarger\sum_{i=1}^{N} z^{(q)}_{ig}n_i\mathbf{x}_i\phi^{(q)}_{ijg}(1-\phi^{(q)}_{ijg})\mathbf{x}_i
	\label{equ:B.13}
\end{equation} 
\begin{equation}
	I(\mathbf{\beta}^{(q)}_{jg}) = \mathlarger\sum_{i=1}^{N} z^{(q)}_{ig}n_i\mathbf{x}'_i v_{ijg}\mathbf{x}_i
	\label{equ:B.14}
\end{equation} 
The updated estimate is 
\begin{equation}
	\beta^{(q+1)}_{jg} = \beta^{(q)}_{jg} + \Bigg(\mathlarger\sum_{i=1}^{N} z^{(q)}_{ig}n_i\mathbf{x}'_i v_{ijg}\mathbf{x}_i\Bigg)^{-1}\Bigg(\mathlarger\sum_{i=1}^{N} z^{(q)}_{ig}\mathbf{x}_i v_{ijg}\zeta^*_{ij}\Bigg)
	\label{equ:B.15}
\end{equation}
\begin{equation}
	\beta^{(q+1)}_{jg} = \Bigg(\mathlarger\sum_{i=1}^{N} z^{(q)}_{ig}n_i\mathbf{x}'_i v_{ijg}\mathbf{x}_i\Bigg)^{-1}\Bigg(\mathlarger\sum_{i=1}^{N} z^{(q)}_{ig}\mathbf{x}_i v_{ijg}\zeta^{(q)}_{ij}\Bigg)
	\label{equ:B.16}
\end{equation}
where $v_{ijg} = \phi^{(q)}_{ijg}(1-\phi^{(q)}_{ijg})$, $\zeta^{(q)}_{ij} = n_i\vec{x}_i\beta^{(q)}_{jg} +\zeta^*_{ij}$ and $\zeta^*_{ij} = y_{ij}/\phi^{(q)}_{ijg} - y_{i1}/\phi^{(q)}_{i1g}$. The weight $v_{ijg}$ and the adjusted response $\zeta^{(q)}_{ij}$ are updated at each iteration based on the current estimates of the multinomial distribution probability $\phi_{ijg}$.	
	\bibliographystyle{apalike}
	\bibliography{mcwm}
\end{document}